\documentclass[numbers,sort&compress]{article}
\usepackage[utf8]{inputenc}
\usepackage{rotating}
\usepackage[normalem]{ulem}
\usepackage{mathtools}
\usepackage{arxiv}
\usepackage{amssymb,amsmath,amsthm}
\usepackage{natbib}
\usepackage{xcolor}
\usepackage{hyperref}
\usepackage[toc,page]{appendix}
\graphicspath{{Artworks/}}
\graphicspath{{imgs/}}
\usepackage{algorithm2e}
\theoremstyle{definition}
\usepackage{url}            %
\usepackage{booktabs}       %
\usepackage{amsfonts}       %
\usepackage{nicefrac}       %
\usepackage{microtype}      %
\usepackage{cleveref}       %
\usepackage{graphicx}
\usepackage{doi}
\author{ \href{https://orcid.org/0000-0002-2393-8056}{\includegraphics[scale=0.06]{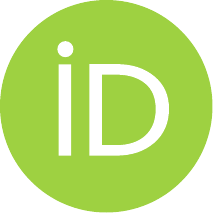}\hspace{1mm}Rohit Goswami} \\
	Science Instiute, Univesity of Iceland \&\\
	Department of Chemistry\\
	Indian Institute of Technology Kanpur \&\\
	Quansight Labs, TX, Austin\\
	\texttt{rgoswami@ieee.org} \\
	\And
	Ashwini Kumar Rawat \\
	Department of Chemistry\\
	Indian Institute of Technology Kanpur\\
	\texttt{ashwinirawat44@gmail.com} \\
	\And
	\href{https://orcid.org/0000-0001-7148-4602}{\includegraphics[scale=0.06]{orcid.pdf}\hspace{1mm}Sonaly Goswami} \\
	Department of Chemistry\\
	Indian Institute of Technology Kanpur\\
	\texttt{sonaly@iitk.ac.in} \\
	\And
	\href{https://orcid.org/0000-0002-2052-0594}{\includegraphics[scale=0.06]{orcid.pdf}\hspace{1mm}Debabrata Goswami}\thanks{Corresponding Author} \\
	Department of Chemistry\\
	Indian Institute of Technology Kanpur\\
	\texttt{dgoswami@iitk.ac.in}
}

\date{\today}
\title{Compositional Analysis of Fragrance Accords Using Femtosecond Thermal Lens Spectroscopy}
\hypersetup{
 pdfauthor={},
 pdftitle={Compositional Analysis of Fragrance Accords Using Femtosecond Thermal Lens Spectroscopy},
 pdfkeywords={},
 pdfsubject={},
 pdfcreator={Emacs 30.1 (Org mode 9.7.29)}, 
 pdflang={English}}
\begin{document}

\maketitle
\maketitle
\begin{abstract} %
Femtosecond thermal lens spectroscopy (FTLS) is a powerful analytical tool, yet
its application to complex, multi-component mixtures like fragrance accords
remains limited. Here, we introduce and validate a unified metric, the
Femtosecond Thermal Lens Integrated Magnitude (FTL-IM), to characterize such
mixtures. The FTL-IM, derived from the integrated signal area, provides a
direct, model-free measure of the total thermo-optical response, including
critical convective effects. Applying the FTL-IM to complex six-component
accords, we demonstrate its utility in predicting a mixture's thermal response
from its composition through linear additivity with respect to component mole
fractions. Our method quantifies the accords' behavior, revealing both the
baseline contributions of components and the dominant, non-linear effects of
highly-active species like Methyl Anthranilate. This consistency is validated
across single-beam Z-scan, dual-beam Z-scan, and time-resolved FTLS
measurements. The metric also demonstrates the necessity of single-beam
measurements for interpreting dual-beam data. This work establishes a rapid,
quantitative method for fragrance analysis, offering advantages for quality
control by directly linking a mixture's bulk thermo-optical properties to its
composition.
\end{abstract}
\keywords{Fragnances, Thermal Lens Spectroscopy, Diffusion, Molecular Interactions, Ultrafast Spectroscopy, Femtosecond Lasers, Optical Properties, Volatile Organic Compounds}
\section{Introduction}
\label{sec:introduction}
Formulating fragrances with desired scent profiles requires precise control over
the composition of complex mixtures of volatile organic compounds.  However,
predicting the behavior of these multi-component mixtures (accords) remains a
significant challenge. Traditional methods for fragrance analysis, such as
sensory panels and gas chromatography-mass spectrometry (GC-MS), can be time-consuming,
expensive, and subjective
\cite{teixeiraPerfumeEngineeringDesign2013,bhattacharyyaPerfumeryMaterialsProduction2009,tollerPerfumeryPsychologyBiology1988a}. Femtosecond thermal lens spectroscopy (FTLS) has been highly effective in
expanding the applicability of Z-scan and time-resolved fixed-point measurements
beyond calculating nonlinear optical material properties \cite{dobekThermalLensingOutside2022,goswamiChapter7Ultrafast2023,d.snookThermalLensSpectrometry1995,chandrasekharTwophotonabsorptionTechniqueSelective2011}. However, its
application to complex, multi-component mixtures, such as fragrance accords,
remains largely unexplored \cite{liuThermalLensSpectrometry2016}. Previous studies
have primarily focused on single-component, binary, or ternary systems
\cite{rawatUnravelingMolecularInteractions2021,mohebbifarStudyThermalBehavior2022,georgesMatrixEffectsThermal2008},
and often struggle to account for convective effects that can significantly
influence measurements in liquids
\cite{kumarUnusualBehaviorThermal2014,singhalThermalLensStudy2019,singhalUnravelingMolecularDependence2020}.
Furthermore, the sensitivity of thermal lensing to molecular-level interactions,
such as hydrogen bonding, suggests its potential for differentiating subtle
differences in mixture composition
\cite{sharmaImpactMolecularConvection2024,rawatPowerdependentStudyPhotothermal2024,rawatInvestigatingPHDependence2023}.
Traditional analysis of FTLS data often relies on fitting analytical models to
experimental data or using summary statistics like the peak-to-valley \(\Delta T_{pv}\) signal
difference or the steady-state thermal lens (SSTL) signal or the signal at \(Z=0\) point
\cite{mauryaEffectFemtosecondLaser2019,sharmaImpactMolecularConvection2024,shenModelCwLaser1992,wangTimeresolvedZscanMeasurements1994}.
These measures can be sensitive to noise, and analytical fits require visual
inspection followed by assumptions about the underlying heat transfer mechanisms
(conduction and convection). We employ a multi-modal FTLS approach
\cite{goswamiIntenseFemtosecondOptical2023} for experimental data acquisition,
focusing on a novel unified measure, the Femtosecond Thermal Lens Integrated Magnitude (FTL-IM).
The FTL-IM is a transformed area metric, defined specifically for each
measurement configuration. It provides a direct, model-independent measure of
the total thermo-optical response, serving as a proxy for the heat-load
dissipation dynamics of these complex systems. This metric accurately accounts
for convective contributions and overcomes limitations of traditional fitting
procedures.

We demonstrate linear additivity of the FTL-IM with respect to component mole
fractions, even in six-component mixtures containing solids, across single-beam
Z-scan, dual-beam Z-scan, and dual-beam time-resolved FTLS measurements,
validating this technique as a simple yet powerful tool for fragrance analysis.
\section{Experimental Section}
\label{sec:expsec}
\subsection{Materials}
\label{sec:orge2ce87c}
Fragrance accords and their individual components (listed in the Supporting
Information) were provided by Jyothy Laboratories Ltd., India, and used as
received.

\begin{figure}[htbp]
\centering
\includegraphics[width=.9\linewidth]{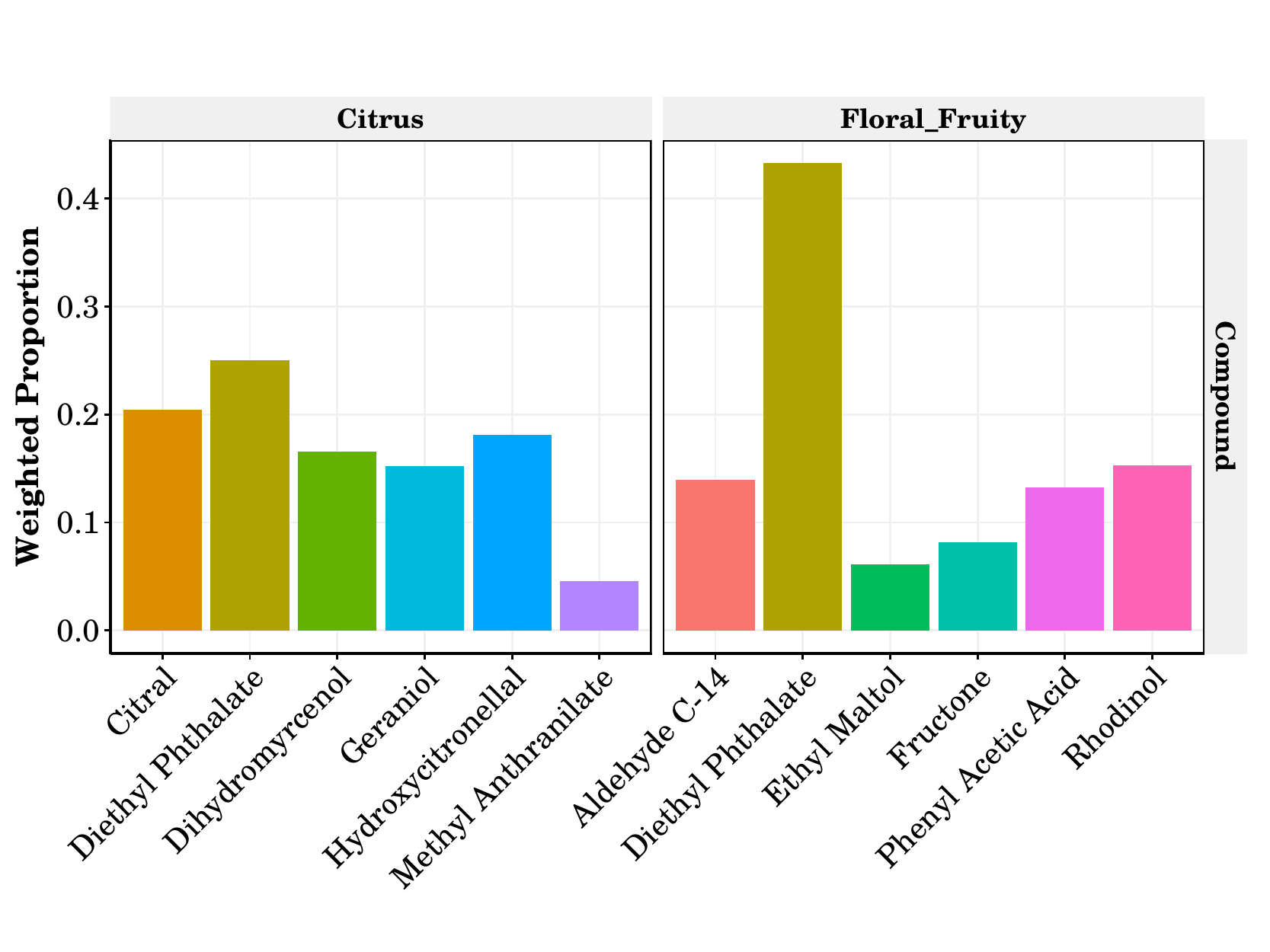}
\caption{\label{fig:mw_comp}Mole fractions of the components in the Citrus (A) and Fruity (B) fragrance accords. Diethyl Phthalate (DEP) is the solvent. Phenyl Acetic Acid and Ethyl Maltol are solids. DEP concentration is higher in the Fruity accord due to the presence of solids. All other compounds are liquids. For exact mole fractions and detailed material properties, please see Tables 1 and 2 in the Supplementary Information.}
\end{figure}

To investigate the thermo-optical properties of complex fragrance mixtures and
test the feasibility of a rapid, quantitative analysis method, we prepared two
fragrance accords, ``Citrus'' and ``Fruity'', with differing compositions (Figure
\ref{fig:mw_comp}). The Citrus accord consisted of six liquid components, while
the Fruity accord consisted of four liquid components and two solid components.
Diethyl Phthalate (DEP) served as the solvent in both accords. Methanol shows a
strong convective effect among the alcohols
\cite{singhalUnravelingMolecularDependence2020} and so was used as a reference.
As shown in Figure \ref{fig:mw_comp}, the Fruity accord has a significantly higher
concentration of DEP, necessitated by the presence of solid components.

\subsection{Thermal Lens Measurements}
\label{sec:org7988234}
Femtosecond thermal lens spectroscopy (FTLS) measurements were performed using
two experimental configurations to obtain three distinct datasets: single-beam
Z-scan, dual-beam Z-scan, and time-resolved pump-probe.

For single-beam Z-scan measurements \cite{singhalUnravelingMolecularDependence2020}, a
mode-locked Ti:Sapphire laser (Coherent Mira-900, \textasciitilde{}150 fs pulse width, 76 MHz
repetition rate, tunable from 730-900 nm) was focused into a 1 mm path length
quartz cuvette containing the fragrance sample.

Dual-beam Z-scan measurements
\cite{kumarImportanceMolecularHeat2014,rawatPowerdependentStudyPhotothermal2024,kumarrawatAchievingMolecularDistinction2022}
employed a mode-mismatched pump-probe configuration using an Er-doped fiber
laser (IMRA Femtolite, 50 MHz repetition rate). A 1560 nm pump beam (\textasciitilde{}300 fs
pulse width) and a 780 nm probe beam (\textasciitilde{}100 fs pulse width) were co-focused into
a 1 mm path length quartz cuvette.

For Z-scan measurements, the sample was translated along the beam propagation
axis (z-direction), and the transmitted light through an aperture was detected
by a silicon photodiode. A 30\% closed aperture was used for single-beam
measurements, and a 40\% open aperture in the far field was used for dual-beam
measurements. The Rayleigh range was determined to be 1.58 mm.

Time-resolved pump-probe measurements
\cite{kumarImportanceMolecularHeat2014,rawatPowerdependentStudyPhotothermal2024,kumarrawatAchievingMolecularDistinction2022}
were performed using the same dual-beam optical setup as above, but with the
sample held at a fixed position, corresponding to the point of maximum thermal
lens (TL) signal (the focus). The relative change in the probe beam signal in
the presence of the pump beam over five-second intervals was collected. Further
details of the experimental setups, including schematic diagrams, are provided
in the Supplementary.

Due to the complex and often proprietary nature of fragrance accord
compositions, a direct quantitative determination of fundamental thermo-optical
parameters (such as \(\frac{dn}{dT}\) or \(n_{2}\)) by analytical fits from the FTLS
data using standard theoretical models is not feasible. However, as we will
demonstrate, the FTL-IM provides a robust and useful metric for characterizing
these mixtures and predicting their behavior, even in the absence of detailed
compositional information.
\section{Computational Methods}
\label{sec:cmeth}
LabVIEW was used for the primary data acquisition \cite{kodoskyLabVIEW2020}.
Subsequent data processing, analysis, and visualization were performed
reproducibly in \texttt{R} \cite{rlang2024} using the \texttt{tidyverse} suite of packages
\cite{wickhamWelcomeTidyverse2019,wickhamGgplot2ElegantGraphics2016}.

To quantify the thermo-optical response from the different FTLS modalities
(single-beam Z-scan, dual-beam Z-scan, and time-resolved pump-probe), we
introduce a unified metric termed the FTLS Integrated Magnitude (FTL-IM). The
FTL-IM is a positive-definite measure of the overall signal magnitude,
calculated by integrating the processed signal as defined by Eq. \ref{eq:ftl_im}.

\begin{equation}
\text{FTL-IM} =
\begin{cases}
|\int_{-\infty}^{\infty} S_{SBZ}(Z) d(Z)| & \text{for single-beam Z-scan} \\
|\int_{-\infty}^{\infty} (1 - S_{DBZ}(Z)) d(Z)| & \text{for dual-beam Z-scan} \\
\frac{1}{N}\int_{0}^{\infty} S_{DTR}(t) dt & \text{for time-resolved}
\end{cases}
\label{eq:ftl_im}
\end{equation}

where:
\begin{itemize}
\item \(S_{SBZ}(z/z_0)\) is the baseline-corrected and Rayleigh-range-normalized
single-beam Z-scan signal.
\item \(S_{DBZ}(z/z_0)\) is the baseline-corrected and normalized dual-beam Z-scan
signal.
\item \(S_{DTR}(t)\) is the baseline-corrected and normalized dual-beam time-resolved
signal.
\item \(Z\) is the normalized z-position, defined by \(z/z_{0}\).
\item \(z_0\) is the Rayleigh range.
\item \(N\) is the number of excitation pulses (N=3 in this study) in the
time-resolved measurements.
\item t is the time.
\end{itemize}

The absolute value is taken over the integrals as the trapezoid rule is used for
calculating the signed area.

For validation and comparison, we contrast our metric with the more traditional
peak-to-valley signal difference for single beam Z-scan data (\(\Delta T_{pv}\))
\cite{sheik-bahaeSensitiveMeasurementOptical1990}, the zero point measure for the dual beam Z-scan data \cite{rawatUnravelingMolecularInteractions2021}, and the steady state TL
signal from time-resolved studies
\cite{shenModelCwLaser1992,sharmaImpactMolecularConvection2024}.

Unlike commonly used point-based measures, the area-based FTL-IM metric is
inherently robust to experimental noise. This is because it relies on area
measurements, avoiding the errors associated with pinpointing exact locations in
noisy data and bypassing the need for analytical models.
\section{Results and Discussion}
\label{sec:results}
\subsection{Single Beam Z-Scan}
\label{sec:orge4f303d}
\begin{figure}[htbp]
\centering
\includegraphics[width=.9\linewidth]{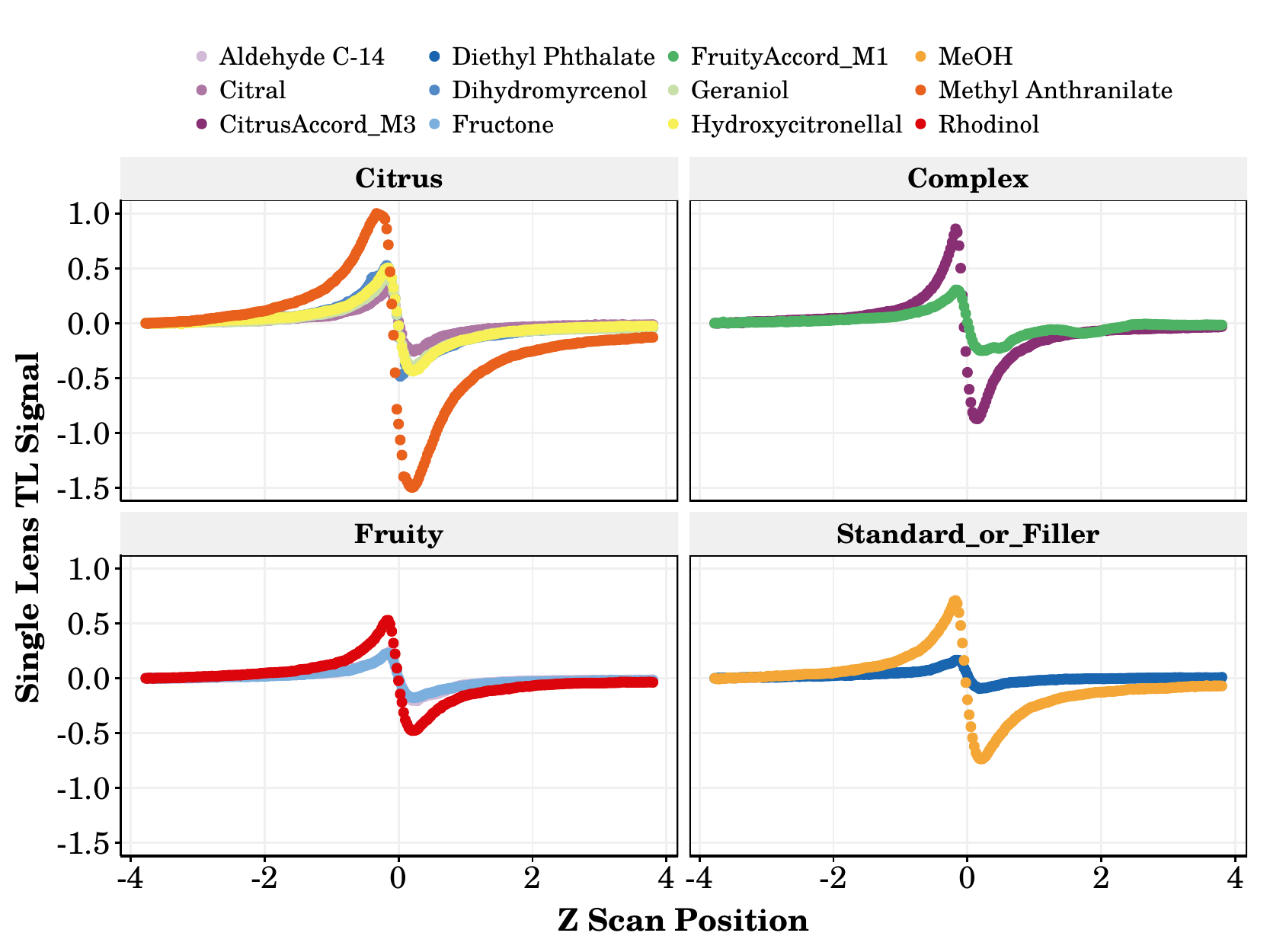}
\caption{\label{fig:sbz_dat}Single beam Z-scan data arranged by accord. MeOH is the reference point, and DEP shows no appreciable signal. Methyl Anthranilate shows an enhanced TL signal which strongly correlates to the final citrus accord signal. The accord signals are in Complex, the components are labeled Fruity or Citrus, and the standard is MeOH, with DEP as the filler (solvent).}
\end{figure}

The single beam data collected is shown in Figure \ref{fig:sbz_dat}, which
demonstrates the prefocal and postfocal transmittance extrema, indicative of a
negative thermo-optic coefficient (\(\frac{dn}{dT}<0\)) for all liquid components
\cite{sheik-bahaeSensitiveMeasurementOptical1990}.

Notably, the Citrus accord curve closely resembles that of Methyl Anthranilate (MeA),
both in shape and magnitude. This suggests that MeA, the only
ester component in the Citrus accord, is the dominant contributor to the overall
heat dissipation dynamics.  The presence of the aromatic ring and carbonyl group
in MeA likely results in a significantly higher absorption
coefficient at the laser wavelength compared to the other, primarily alcohol and
aldehyde, components. Furthermore, MeA and aldehydes, like Hydroxycitronellal, could react to form a Schiff base \cite{arctanderPerfumeFlavorChemicals2017}. This reaction may alter the absorption properties and contribute to the enhanced thermal lens signal observed in the Citrus accord.
The Fruity accord, in contrast,
exhibits a smaller signal magnitude, consistent with the higher concentration of
DEP (which shows negligible signal) and the presence of solid components.

Furthermore, a closer examination of the individual component curves (Figure S3)
reveals a striking similarity between the Z-scan curves of Rhodinol, Geraniol,
and Hydroxycitronellal. Chemically, this is not surprising since Rhodinol is a
mixture of Geraniol and Citronellol \cite{arctanderPerfumeFlavorChemicals2017}.
From a perfumery perspective, however, this similarity is noteworthy. These
compounds, all contributing to floral scent notes, despite their differing
molecular weights (see Supplementary Table 2), exhibit nearly indistinguishable
thermal dissipation dynamics in the single-beam Z-scan. This suggests that, from
the perspective of the thermal lens signal, these components could be considered
somewhat interchangeable within a formulation, and indeed, these three,
Rhodinol, Geraniol, and Hydroxycitronellal, have ``Floral'' scents.

\begin{figure}[htbp]
\centering
\includegraphics[width=.9\linewidth]{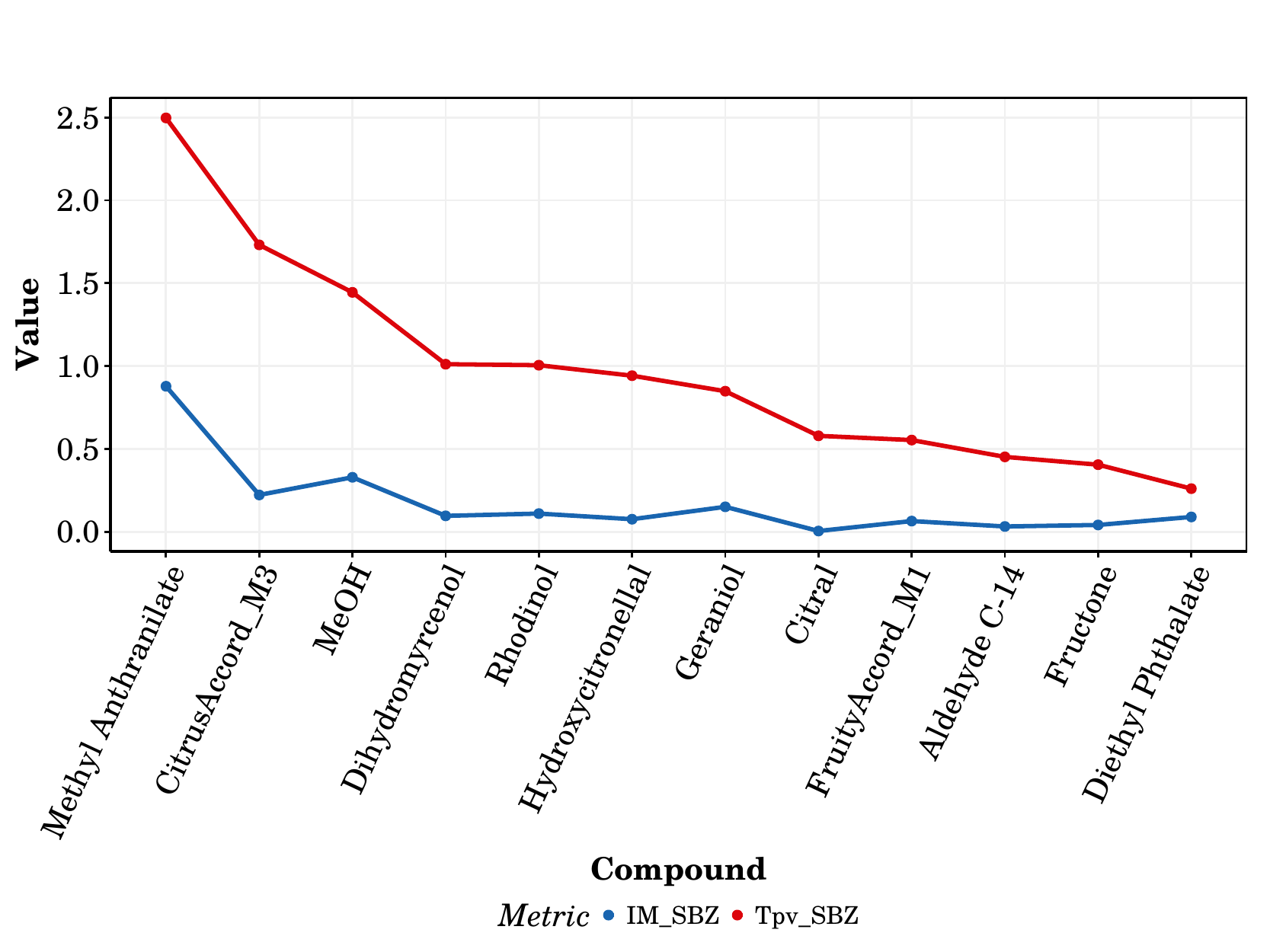}
\caption{\label{fig:ftls_im_tpv}FTL-IM and peak-to-valley (\(\Delta T_{pv}\)) measures for single beam Z-scan data. Most trends are similar for both measures; however, \(\Delta T_{pv}\) is qualitatively incorrect for the data observed, as discussed in the text, as MeOH should be higher than Citrus, and Geraniol should be higher than Hydroxycitronellal. Additionally, Citral should have a lower signal than the Fruity accord. Fructone, DEP, Aldehyde C-14 and Citral show no appreciable signal.}
\end{figure}

Our chemical and visual observations are reinforced by the FTL-IM summary
statistic as seen in Figure \ref{fig:ftls_im_tpv}, which also includes a
comparison against the peak-to-valley transmittance difference (\(\Delta
T_{pv}\)). We note that the broadening of the Z-scan signal is captured
accurately by the FTL-IM, which in turn correctly orders the signal strength and
thermal dissipation of MeOH and the Citrus Accord. Hydroxycitronellal has a
higher molecular weight compared to Geraniol, so it is expected to have slightly
slower dynamics than the lighter Geraniol, a relation which is seen in the
FTL-IM but not in the \(\Delta T_{pv}\).

Citral and Fruity accord have similar \(\Delta T_{pv}\) values, although the
Fruity accord has a larger FTL-IM measure due to its shape. This is in line with
the trends shown for dual beam data, which suggests that the measure is
correctly proxying the thermal dissipation dynamics of the system. In terms of
magnitude, the absolute value of the \(\Delta T_{pv}\) value is higher, as the
signed area is used for FTL-IM, though the \(\Delta T_{pv}\) is unable to
demonstrate sensitivity to convection and other thermal heat dissipation modes.
\subsection{Dual Beam}
\label{sec:org8973396}
\begin{figure}[htbp]
\centering
\includegraphics[width=.9\linewidth]{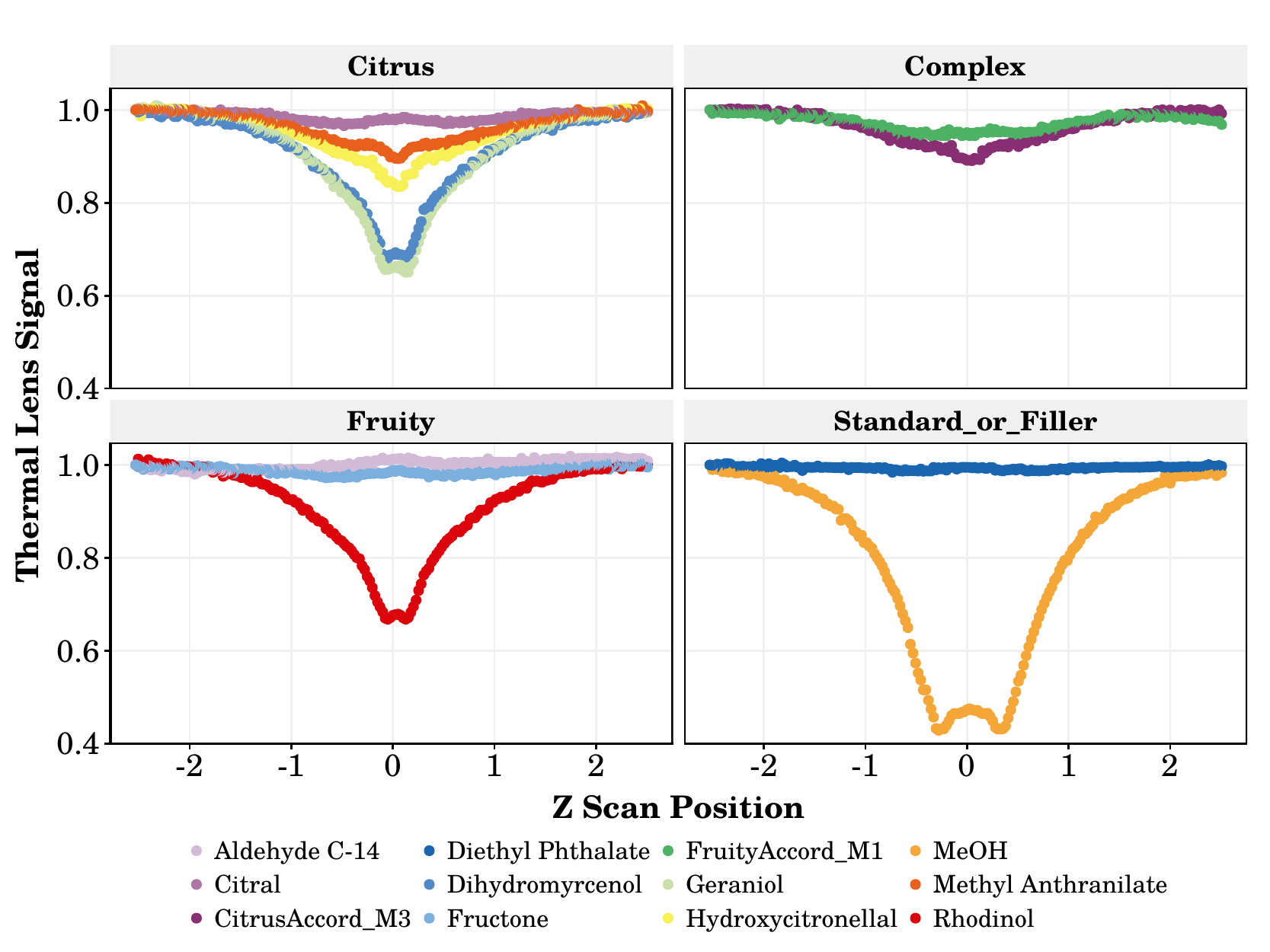}
\caption{\label{fig:dbz_dat}Dual-beam Z-scan data arranged by accord. MeOH is the reference point, and DEP shows no appreciable signal. Methyl Anthranilate does not show an enhanced TL signal though the final citrus accord signal matches the shape very closely. Rhodinol, Dihydromyrcenol, and Geraniol have similar signals. Fructone, DEP, Aldehyde C-14 and Citral show no appreciable signal.}
\end{figure}

Dual-beam measurements were performed to further investigate the TL signal of
the individual components and accords. The data collected from the dual beam
Z-scan is shown in Figure \ref{fig:dbz_dat}. The alcohol curves: Methanol,
Rhodinol, Dihydromyrcenol, and Geraniol exhibit the characteristic ``W'' shape
expected for materials which are known to dissipate heat through convection.
Diethyl Phthalate, Fructone, Aldehyde C-14, and Citral, are completely colorless
and show no appreciable signal. The Citrus accord curve is almost exactly
overlapped with that of MeA, which may be understood in the context of the
single beam signal in the previous section.

\begin{figure}[htbp]
\centering
\includegraphics[width=.9\linewidth]{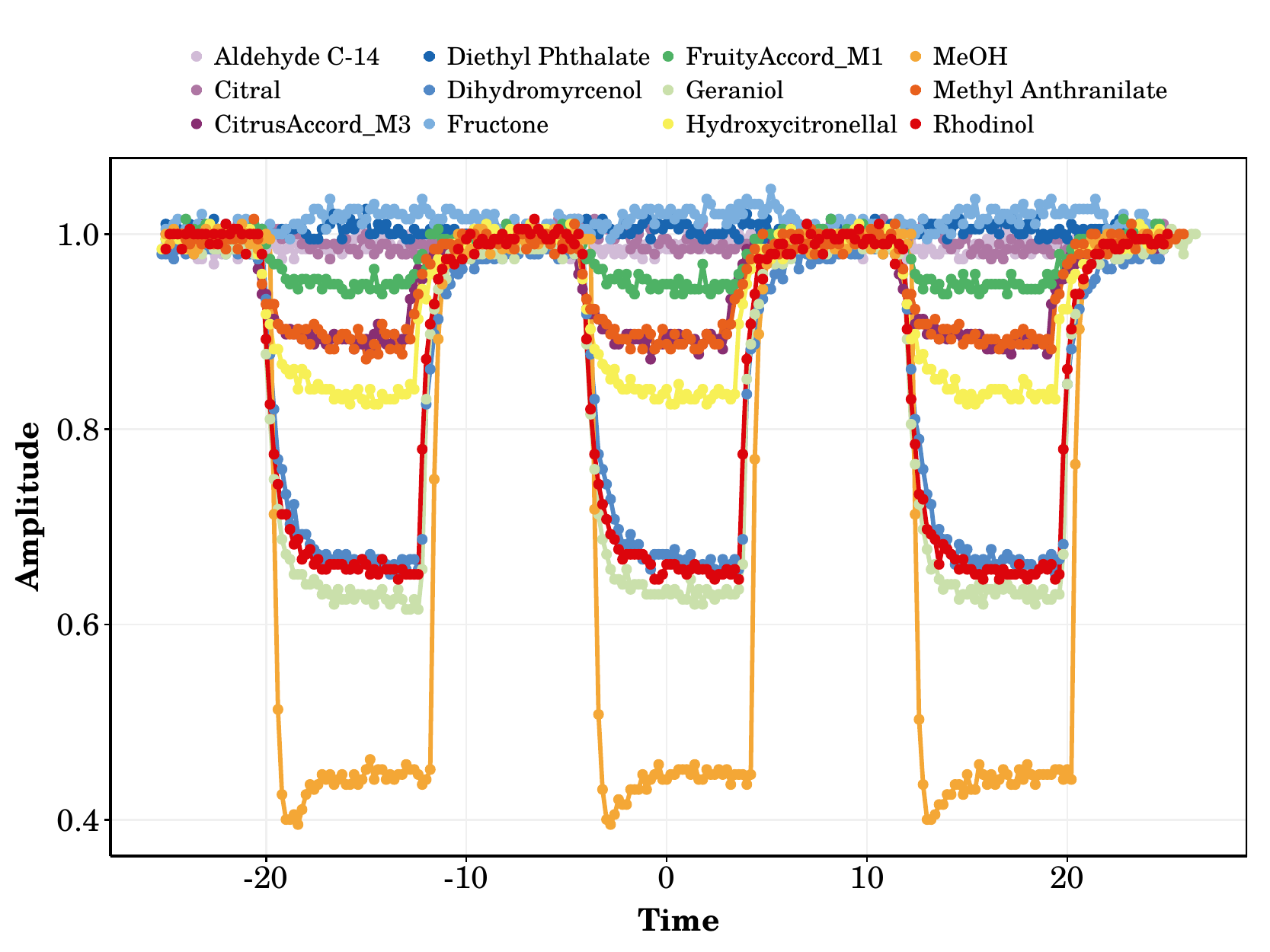}
\caption{\label{fig:dtr_dat}Dual-beam time-resolved data arranged by accord. MeOH is the reference point, and Fructone, DEP, Aldehyde C-14, and Citral show no appreciable signal.}
\end{figure}

The pump-probe time-resolved experimental data is shown in Figure
\ref{fig:dtr_dat} and reinforces the same trends and concepts, with the dip below
the steady-state signal indicating heat dissipation beyond conduction.

\subsection{FTL-IM Consistency and Linearity}
\label{sec:org2a38a19}

\begin{figure}[htbp]
\centering
\includegraphics[width=.9\linewidth]{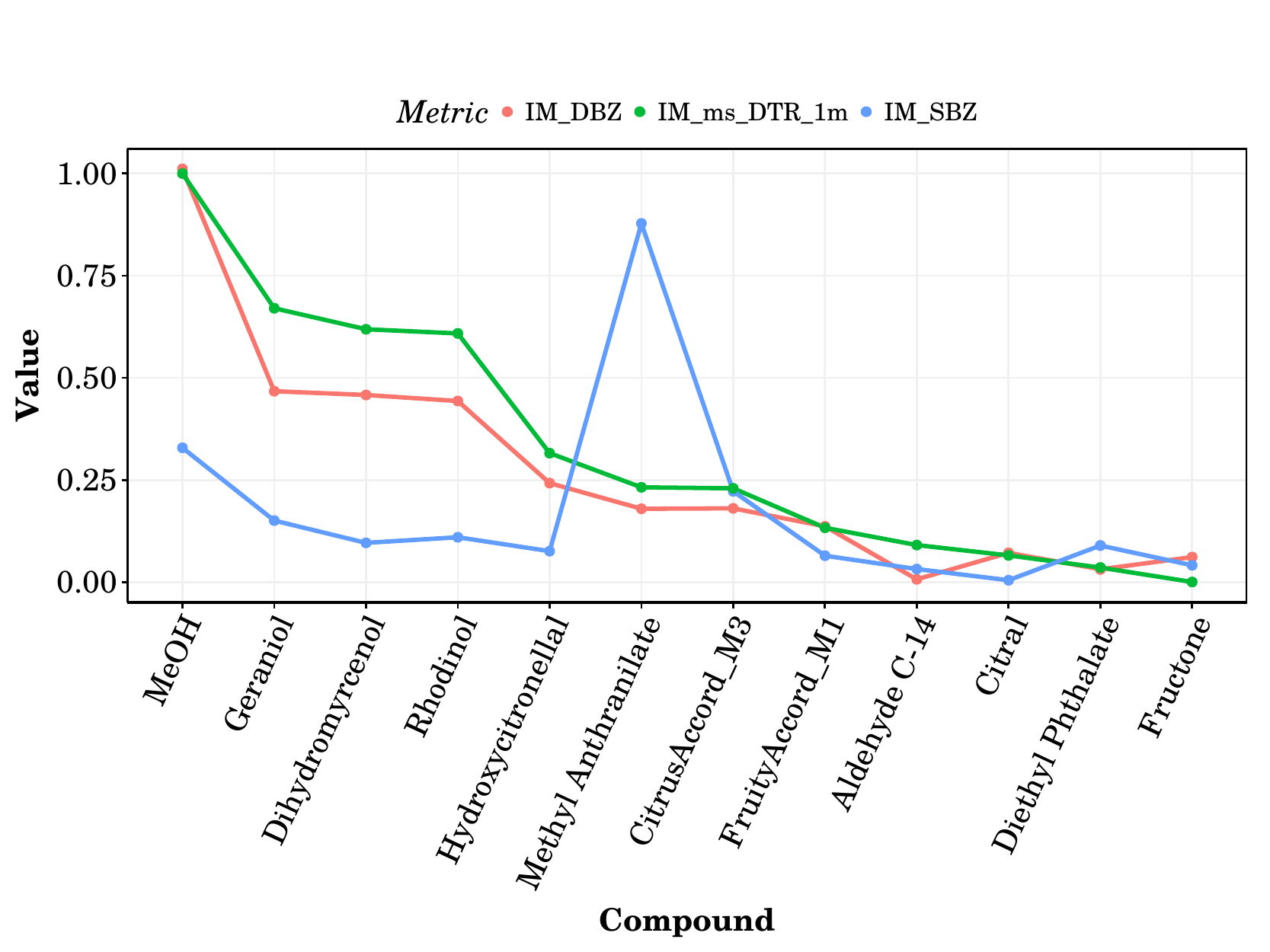}
\caption{\label{fig:ftls_im_comparison}Comparison of FTL-IM metrics, after normalization for the FTL-IM for dual beam time-resolved data, demonstrating the remarkable similarity in trends for the measurements of all three dual beam Z-scan data, along with the MeA (ester) spike in the single beam data. Fructone, DEP, Aldehyde C-14, and Citral show no appreciable signal. Values for each modality have been scaled for visual comparison of trends (see Supplementary for details).}
\end{figure}

FTLS Integrated Magnitude (FTLS-IM) values for individual components and
accords, measured using single-beam Z-scan (IM\(_{SBZ}\)), dual-beam Z-scan
(IM\(_{DBZ}\)), and time-resolved pump-probe (IM\(_{DTR}\)) are presented in
Figure \ref{fig:ftls_im_comparison}.

Despite the differences in experimental configuration for each modality, the
overall trends observed in Figure \ref{fig:ftls_im_comparison} are remarkably
consistent. This consistency arises from the fact that all three techniques are
fundamentally probing the same underlying phenomenon: the power-dependent
heating of the sample by the laser beam and the resulting change in refractive
index. While the single-beam Z-scan is primarily sensitive to the spatial
profile of the induced thermal lens, the dual-beam Z-scan is more directly
sensitive to the phase shift and convective effects, and the time-resolved
measurements probe the temporal dynamics of the thermal lens formation and
decay, they all ultimately depend on the amount of energy absorbed by the sample
and converted to heat. Crucially, we have demonstrated in the previous
sub-section that the measure is able to account for convective and conductive
heating effects, and is able to provide a quantitative view of processes taking
place in the ester, which has not yet been analytically modeled.

The observed consistency justifies our use of a unified FTL-IM metric as defined
in Eq. \ref{eq:ftl_im}. It also highlights the importance of the single-beam
measurements, which provide a sensitive measure of the overall signal strength
and the underlying dissipation dynamics. As a check for linearity in FTL-IM, we
predict the values of the accords using a mole-fraction weighted average, using
Eq. \ref{eq:ftl_im_pred}.

\begin{equation}
\label{eq:ftl_im_pred}
IM_{wAvg} = \sum_{i} (IM_i \cdot MFrac_i)
\end{equation}

where \(IM_{i}\) represents the measured IM for each component, and \(MFrac_{i}\) is
the corresponding mole fraction. As shown in Table \ref{tbl:pred_ftlim_fruity} and
Table \ref{tbl:pred_ftlim_citrus}, the FTL-IM correlates linearly with component
mole fractions. The relatively larger errors in the single beam (IM\textsubscript{SBZ})
Citrus accord prediction arises from the down-weighting of MeA, which, despite
its low mole fraction, significantly influences the single-beam Z-scan curve
shape. For the fruity accord, the linearity assumption holds much better as no
single component dominates the response.

\begin{table}[htbp]
\caption{\label{tbl:pred_ftlim_fruity}Predicted FTL-IM for the Fruity accord based on the weighted average of mole fraction for each FTLS modality.}
\centering
\begin{tabular}{lrrrl}
Measure & Pred. & Exp. & Abs. Err. & Obs. Range\\
\hline
\(IM_{DBZ}\) & 0.11 & 0.14 & 0.03 & (0.07, 0.47)\\
\(IM_{DTR}\) & 16.18 & 16.26 & 0.08 & (13.69, 16.59)\\
\(IM_{SBZ}\) & 0.08 & 0.06 & 0.02 & (0.0, 0.88)\\
\end{tabular}
\end{table}

\begin{table}[htbp]
\caption{\label{tbl:pred_ftlim_citrus}Predicted FTL-IM for the Citrus accord based on weighted average of mole fraction for each FTLS modality.}
\centering
\begin{tabular}{lrrrl}
Measure & Pred. & Exp. & Abs. Err. & Obs. Range\\
\hline
\(IM_{DBZ}\) & 0.22 & 0.18 & 0.04 & (0.01, 0.44)\\
\(IM_{DTR}\) & 15.49 & 15.8 & 0.31 & (13.98, 16.90)\\
\(IM_{SBZ}\) & 0.12 & 0.22 & 0.1 & (0.03, 0.11)\\
\end{tabular}
\end{table}
\section{Conclusion}
\label{sec:conclusion}
We have demonstrated linear additivity between the femtosecond thermal lens
spectroscopy Integrated Magnitude (FTL-IM) and component mole fractions in
complex, multi-component fragrance accords. The FTL-IM, defined as the absolute
value of the signed integral of a baseline-corrected, Rayleigh-range-normalized
signal for single-beam Z-scan, and as the area under the baseline-corrected,
normalized curve for dual-beam Z-scan and time-resolved measurements, provides a
unified metric for characterizing the underlying thermal dissipation dynamics
across these modalities. This linearity, observed for both Citrus and Fruity
accords across all three FTLS techniques, demonstrates the robustness of the
FTL-IM for analyzing complex mixtures, even when including solid components in
unknown compositions.

As a reactive and volatile ester, Methyl Anthranilate (MeA) shows an outsized
signal, going beyond the reference MeOH. From the single beam data, both from
visual inspection and quantified by FTL-IM or even \(\Delta T_{pv}\), it is clear
that the presence of MeA dominates the overall heat dissipation mechanism for
the citrus accord, leading to a signal similar in shape to MeA, despite having a
relatively low presence in terms of weight fractions.

Our results underscore the value of performing single-beam Z-scan studies, especially for volatile and reactive esters such as MeA, while highlighting the similarity of information obtained from dual-beam time-resolved and Z-scan studies (though time-resolved data has a stronger signal). The rapid,
non-destructive nature of FTLS, combined with the predictive power of the
FTL-IM, offers significant advantages for fragrance quality control and
formulation. While further research is needed to establish a direct correlation
between the FTLS signal and olfactory perception, our results hint at a
potential link between readily measurable thermo-optical properties and the
ultimate scent profile of a fragrance mixture. The characterization of complex chemical mixtures is also a foundational challenge in fields ranging from materials science to industrial product formulation \cite{sadiqCriticalReviewMetalorganic2024}. Future work will explore a wider
range of mixtures, perform rigorous statistical validation of the model, and investigate potential correlations between FTL-IM and olfactory
properties, heat, and mass transport, along with classical thermo-optic
properties.
\section{Acknowledgments}
\label{sec:org4552b46}
R.G. thanks Mrs. Ruhila Goswami for fruitful discussions. R.G. and A.K.R. thank
the Indian Institute of Technology Kanpur (India) for their respective Research
Fellowship and Senior Research Fellowship grants. R.G. was also partially
supported by the Icelandic Research Fund (grant no. 217436-053). D.G. thanks
funding support from the SERB Core Research Grant, Govt. of India. We dedicate
this work to the memory of our beloved little birdie, Tuitui.
\section*{Conflict of interest}
\label{sec:orgce80387}
The authors declare no conflict of interest.

\begin{appendices}
\vspace{0.7em}
\noindent \textbf{Exact code to reproduce all models and visualization is available on reasonable request.}
\renewcommand{\thepage}{S\arabic{page}}
\renewcommand{\thesection}{S\arabic{section}}
\renewcommand{\thetable}{S\arabic{table}}
\renewcommand{\thefigure}{S\arabic{figure}}
\setcounter{figure}{0}
\section{Experimental}
\label{sec:experimental}
\subsection{Materials used}
\label{sec:org81a0e45}
The following materials were analyzed in this work: Citral, Fractone, Diethyl
phthalate, Geraniol, Rhodinol, C4 Aldehyde, Dihydro Mercenol, Hydroxy
Citronellal, Methyl Anthranilate, and the fragrance accords ``Citrus'' and
``Fruity.''  These materials, provided to us by Jyothy Laboratories Ltd, India, were
characterized using single-beam and dual-beam femtosecond thermal lens
spectroscopy (FTLS). Basic material properties of interest are listed in Table
\ref{tbl:mprop} and Table \ref{tbl:addl}, with abbreviations defined in Table \ref{tbl:addl_abbrev}.

\begin{table}[htbp]
\caption{\label{tbl:mprop}Material properties of compounds in the study. Abbreviations are in Table \ref{tbl:addl_abbrev}.}
\centering
\begin{tabular}{lrlrl}
Compound & CAS-ID & Accord & M. Frac & Appearance\\
\hline
Rhodinol & 141-25-3 & Floral-Fruity & 15 & Colorless / Pale yellow clear liquid\\
Phenyl Acetic Acid & 103-82-2 & Floral-Fruity & 13 & White crystalline powder\\
Ethyl Maltol & 4940-11-8 & Floral-Fruity & 6 & White crystalline\\
Aldehyde C-14 & 104-67-6 & Floral-Fruity & 13.70 & Colorless / Pale yellow clear liquid\\
Fructone & 6413-10-1 & Floral-Fruity & 8 & Colorless clear liquid\\
Citral & 5392-40-5 & Citrus & 22.20 & Colorless / Pale yellow clear liquid\\
Dihydromyrcenol & 18479-58-8 & Citrus & 18 & Colorless clear liquid\\
Hydroxycitronellal & 107-75-5 & Citrus & 19.70 & Pale yellow thick liquid\\
Geraniol & 106-24-1 & Citrus & 16.50 & Colorless clear liquid\\
Methyl Anthranilate & 134-20-3 & Citrus & 5 & Pale / Dark yellow liquid\\
Diethyl Phthalate & 84-66-2 & Citrus & 27.19 & Colorless liquid\\
Diethyl Phthalate & 84-66-2 & Floral-Fruity & 42.50 & Colorless liquid\\
\end{tabular}
\end{table}

\begin{table}[htbp]
\caption{\label{tbl:addl}Additional scent related properties of compounds in the study. Abbreviations in Table \ref{tbl:addl_abbrev}.}
\centering
\begin{tabular}{lrrllllrl}
Compound & MolWt & logPow & OType & OStr & Volatility & SType & S\textsubscript{hr} & Nature\\
\hline
Rhodinol & 156.27 & 3.24 & Floral & Med. & Heart & Rose & 236 & Alcohol\\
Phenyl Acetic Acid & 136.15 & 1.42 & Honey & High & Heart & Honey & 400 & Acid\\
Ethyl Maltol & 140.14 & 0.63 & Caramel\textsubscript{Vanilla} & High & Base & Fruity & 360 & Alcohol\\
Aldehyde C-14 & 206.24 & 1.65 & Fruity & Med. & Base & Peach & 338 & Aldehyde\\
Fructone & 174.19 & 0.98 & Fruity & Med. & HH & Apple & 12 & Ketal\\
Citral & 152.24 & 3.45 & Citrus & Med. & Head & Lemon & 12 & Aldehyde\\
Dihydromyrcenol & 156.27 & 2.99 & Citrus & Med. & Heart & Fresh & 16 & Alcohol\\
Hydroxycitronellal & 172.27 & 1.65 & Floral & Med. & Heart & Muguet & 218 & Aldehyde\\
Geraniol & 154.25 & 3.56 & Floral & Med. & HH & Rose & 60 & Alcohol\\
Methyl Anthranilate & 151.17 & 1.88 & Fruity & Med. & Heart & Neroli & 88 & Ester\\
Diethyl Phthalate & 222.24 & 2.42 &  &  & Solv. &  & 400 & Alcohol\\
Diethyl Phthalate & 222.24 & 2.42 &  &  & Solv. &  & 400 & Alcohol\\
\end{tabular}
\end{table}

\begin{table}[htbp]
\caption{\label{tbl:addl_abbrev}Abbreviations used in Table \ref{tbl:mprop} and Table \ref{tbl:addl}.}
\centering
\begin{tabular}{ll}
Abbreviation & Term\\
M.Frac. & Mole Fraction\\
CAS-ID & Chemical abstracts service number\\
MolWt & Molecular Weight\\
logPow & Logarithm (base 10) of the Octanol-water partition coefficient\\
OType & Odor type\\
OStr & Odor strength\\
SType & Scent Type\\
S\textsubscript{hr} & Substantivity in hours (at 100\%)\\
HH & Head or heart\\
Solv. & Solvent\\
\end{tabular}
\end{table}
\subsection{Experimental Setup}
\label{sec:org06aa0e4}
LabVIEW \cite{kodoskyLabVIEW2020} is used to automate the instruments and to record data.
\subsubsection{Single Beam Z-Scan}
\label{sec:orgd697d3b}
\begin{figure}[htbp]
\centering
\includegraphics[width=.9\linewidth]{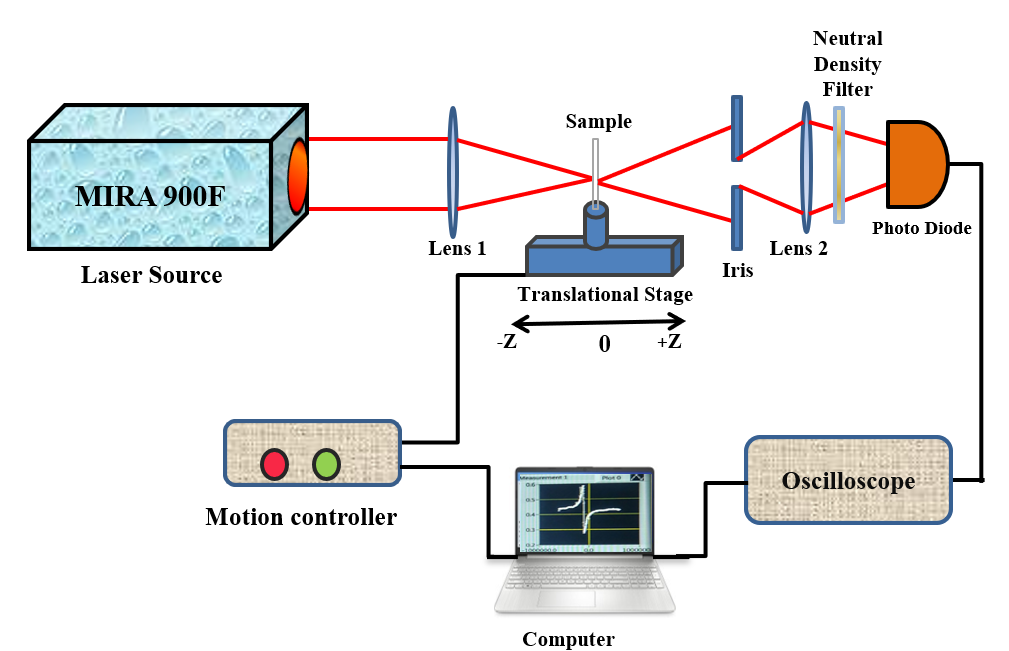}
\caption{\label{fig:cazs}Schematic experimental setup for the close aperture z-scan.}
\end{figure}

For the single-beam setup, a mode-locked Ti:Sapphire laser (Coherent Mira-900,
100-200 fs pulse width, 730-900 nm tuning range, 76 MHz repetition rate) was
focused using a 20 cm focal length lens into a 1 mm path length quartz cuvette
containing the fragrance sample. Sample translation along the beam axis
(z-direction) was controlled by a motorized stage (Newport, 0.1 µm resolution).
Transmitted light through a 30\% closed aperture was detected by a silicon
photodiode (Thorlabs PDA100A-EC). The average laser power was 350 mW. Further
details of the close-aperture single beam z-scan technique are described elsewhere
\cite{singhalUnravelingMolecularDependence2020} and a schematic is given in Figure \ref{fig:cazs}.

\subsubsection{Dual Beam}
\label{sec:org10cd92a}
\begin{figure}[htbp]
\centering
\includegraphics[width=.9\linewidth]{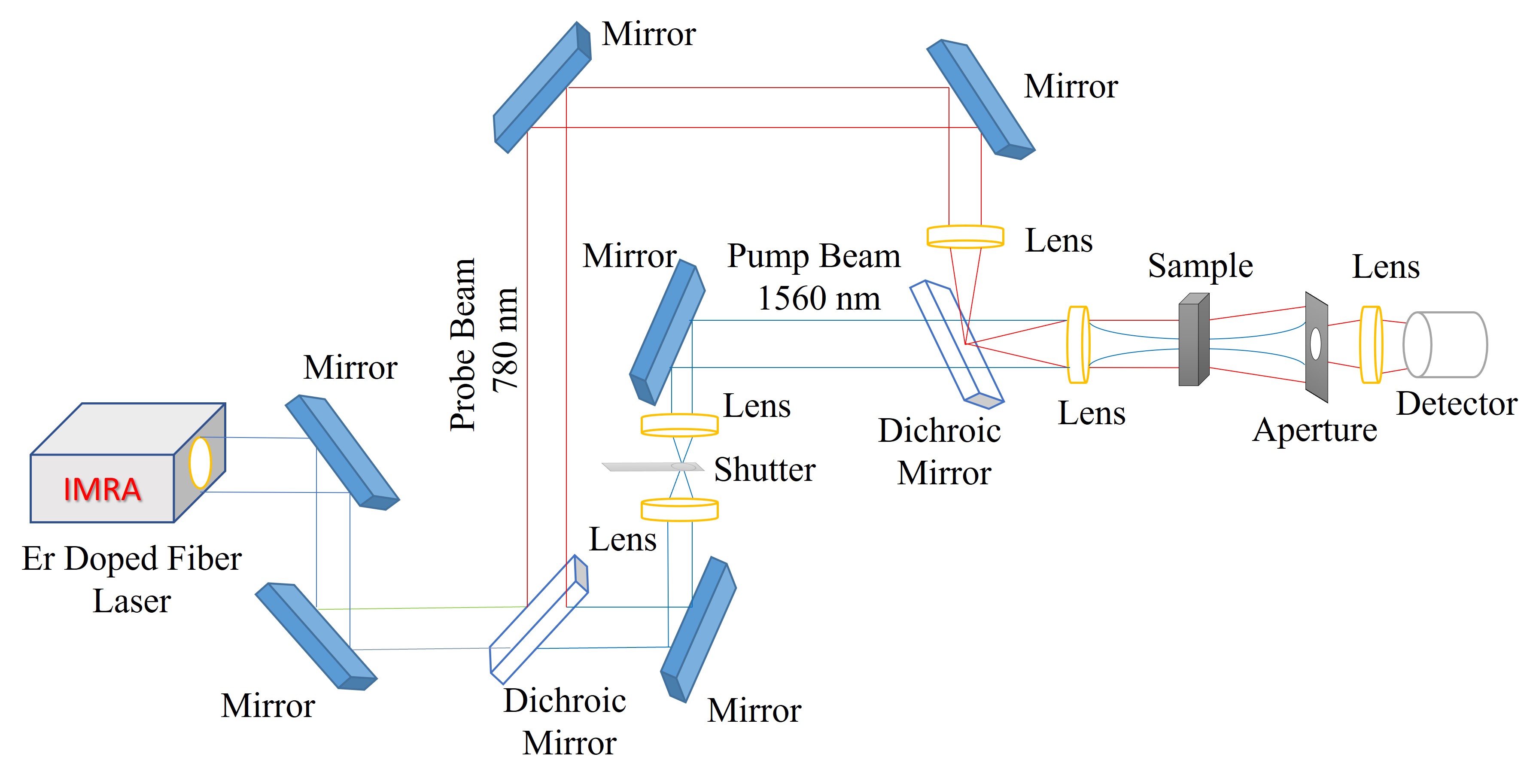}
\caption{\label{fig:dbexp}A schematic diagram for a dual-beam experimental setup.}
\end{figure}

A mode-mismatched pump-probe configuration \cite{shenModelCwLaser1992} was
employed for the dual-beam setup
\cite{kumarUnusualBehaviorThermal2014,rawatPowerdependentStudyPhotothermal2024,kumarrawatAchievingMolecularDistinction2022}.
An Er-doped fiber laser (IMRA Femtolite, 50 MHz repetition rate) provided a 1560
nm pump beam (300 fs pulse width, 10 mW average power) and a 780 nm probe beam
(Gaussian \(TEM_{00}\), 100 fs pulse width, 6 mW mean power). The distance between the
sample and aperture is maintained to satisfy the far field diffraction limits
and a cut-off filter is positioned to block the pump before the detector. An
oscilloscope (LeCroy Wave Runner 64xi, 600 MHz) is connected to the photodiode
and controlled via LabVIEW. A schematic for this setup is shown in Figure \ref{fig:dbexp}.

\begin{enumerate}
\item Z-Scan
\label{sec:org5acd1eb}
The probe beam was focused using a 5 cm focal length plano-convex lens into the
sample cuvette, and the transmitted light through a 60\% closed aperture was
detected by a silicon photodiode (Thorlabs PDA100A-EC).
\item Pump-probe
\label{sec:org7204cfe}
A mechanical shutter is added to the pump arm with an activation time of less
than 500 \(\mu\) s. The shutter is opened and closed several times with fixed open
and shut times until the TL (thermal lens) steady state is achieved
\cite{kumarImportanceMolecularHeat2014,kumarrawatAchievingMolecularDistinction2022}.
\end{enumerate}
\section{Computational details}
\label{sec:org6e52b99}
\subsection{Area Calculation}
\label{sec:org1fb8534}
The integration is performed numerically using the trapezoidal rule
\cite{eppersonIntroductionNumericalMethods2012}. This approximates the definite
integral of a function by dividing the area under the curve into a series of
trapezoids and summing their areas.  Given a set of \(n\) data points \((x_i,
y_i)\), where \(i = 1, 2, ..., n\), the area, \(A\), is approximated as:

\begin{equation}
A = \frac{1}{2} \sum_{i=1}^{n-1} (x_{i+1} - x_i)(y_i + y_{i+1})
\label{eq:trapezoidal_rule}
\end{equation}

This numerical integration method is equivalent to the \texttt{trapz} function in the
\texttt{pracma} package \cite{pracma2023}, which, in turn, is based on the Gauss
polygon area formula.  For a closed polygon with \(n\) vertices \((x_i, y_i)\), the
Gauss polygon area formula is given by:

\begin{equation}
A = \frac{1}{2} \left| \sum_{i=1}^{n} (x_i y_{i+1} - x_{i+1} y_i) \right|
\label{eq:gauss_area}
\end{equation}

where \(x_{n+1} = x_1\) and \(y_{n+1} = y_1\) (i.e., the polygon is implicitly
closed) \cite{allgowerComputingVolumesPolyhedra1986}. In the context of our Z-scan
and time-resolved data, the \(x_i\) values correspond to sample position (Z-scan)
or time (time-resolved), and the \(y_i\) values correspond to the normalized
transmittance. Areas calculated where the curve travels to the left are counted
negatively, and areas to the right are positive; the total area is the sum of
all such areas, taking sign into account.
\subsection{Single Beam Z-Scan FTLS-IM}
\label{sec:org425b4ab}
For single-beam Z-scan measurements, the FTLS-IM is defined as the integral of
the absolute value of the baseline-corrected and Rayleigh-range-normalized
Z-scan signal:

\begin{equation}
IM_{SBZ} = |\int_{-\infty}^{\infty} |S(z/z_0)| d(z/z_0)|
\label{eq:im_sb}
\end{equation}

where  \(S(z/z_0)\) is the baseline-corrected Z-scan signal as a function of the
normalized sample position (\(z/z_0\)), and \(z_0\) is the Rayleigh range. The
baseline correction involves subtracting the average of the signal values far
from the focus (\(z/z_0\) \(<< 0\) and \(z/z_0\) \(>> 0\)) from the raw signal.

\begin{description}
\item[{Baseline Correction}] For each Z-scan curve, a baseline value was determined
by the first recorded thermal lensing signal value at the lowest normalized
Z-position. This initial value was then subtracted from all subsequent thermal
lensing signal measurements for that specific compound to remove any
background signal.
\item[{Rayleigh Range Normalization}] The sample position (ZPos) was normalized by a
factor related to the Rayleigh range. With a Rayleigh range \(z_0 = 1.58\) mm,
the dimensionless z-position (\(ZPos_{norm}\)) was obtained by dividing the ZPos
by \(1.58 \times 10^5\) mm (since \(z_0 \times 10^5\) was used as the
normalization factor).
\end{description}
\subsubsection{Preprocessing}
\label{sec:org9b328a5}
The initial data frame was prepared by converting the compound identifier to a
factor variable and normalizing the Z-position values. The normalization was
achieved by dividing the Z-position by \(1.58 \times 10^5\) (assuming a Rayleigh range of 1.58 mm).

For each compound, a baseline correction was applied to the thermal lensing
signal. This involved identifying the first recorded thermal lensing signal
value at the lowest normalized Z-position and subtracting this initial value
from all subsequent thermal lensing signal measurements for that specific
compound. This step ensured that the initial signal was set to zero for each
compound.

Following the baseline correction, the thermal lensing signal data for each
compound was normalized. This normalization was performed by dividing all
baseline-corrected signal values by the maximum absolute baseline-corrected
signal value observed across all compounds. This step scaled the data to a range
between -1 and 1.
\subsection{Dual Beam Z-Scan FTLS-IM}
\label{sec:orgc16a926}
For dual-beam Z-scan measurements, the FTLS-IM is defined as the area under the
curve (AUC) of the normalized and baseline-corrected Z-scan signal:

\begin{equation}
IM_{DB} = |\int_{-\infty}^{\infty} 1 - S(z/z_0) d(z/z_0)|
\label{eq:im_db}
\end{equation}

where \(S(z/z_0)\) is the baseline-corrected and normalized Z-scan signal as a
function of the normalized sample position (\(z/z_0\)), and  \(z_0\) is the Rayleigh
range. The baseline correction involves subtracting 1 from the Z-scan signal,
which has been normalized so that the initial value is unity.
\subsubsection{Preprocessing}
\label{sec:orgac45dfc}
The initial data frame was prepared by converting the compound identifier to a
factor variable and normalizing the Z-position values. The normalization was
achieved by dividing the Z-position by \(1.58 \times 10^5\) (assuming a Rayleigh range of 1.58 mm).

Following this, for each compound, a baseline correction was applied
to the thermal lensing signal. The first recorded thermal lensing signal value
was identified (assuming the data was ordered by the Z-position). Then, all
thermal lensing signal values for that compound were shifted by subtracting this
first value and adding 1. This step ensured that the initial thermal lensing
signal value for each compound was approximately unity.
\subsection{Dual Beam Time Resolved FTLS-IM}
\label{sec:orgee8654c}
For time-resolved pump-probe measurements, the FTLS-IM is defined as the area
under the curve (AUC) of the normalized and baseline-corrected time-dependent
signal:

\begin{equation}
IM_{TR} = \frac{1}{N}\int_{0}^{\infty} S_{DTR}(t) dt
\label{eq:im_dtr}
\end{equation}

where \(S(t)\) is the time-resolved signal, \(t\) is the total time, and \(N\) is the
number of excitation pulses (N=3 in this study). The signal is normalized by
applying the Pruned Exact Linear Time (PELT) algorithm with the Modified Bayes
Information Criterion (MBIC) to identify the initial response, setting this
value to unity. Subsequently, the signal was shifted such that the minimum
amplitude value was zero. This normalization allows for a direct comparison of
the signal decay across different samples, accounting for variations in initial
signal amplitude and baseline offsets.
\subsubsection{Thinning}
\label{sec:org4b496f1}
To reduce computational cost and data volume, the time-resolved data were
subsampled, retaining every 100th data point. This subsampling rate was chosen
to be well above the Nyquist frequency, ensuring that all relevant features of
the thermal lens signal were preserved.
\subsubsection{Normalization}
\label{sec:org2a3e665}
Preprocessing and normalization of the time resolved data involved the Pruned
Exact Linear Time (PELT) algorithm \cite{killickOptimalDetectionChangepoints2012}
with the Modified Bayes Information Criterion (MBIC)
\cite{zhangModifiedBayesInformation2007} penalty as implemented in the
\texttt{changepoint} package \cite{killickChangepointPackageChangepoint2014}.

For each compound, this algorithm was used to detect the first significant
change in the mean of the amplitude signal. The amplitude value at this detected
changepoint (or the first amplitude value if no changepoint was identified) was
then used to normalize the signal by subtracting this value and adding 1. This
effectively sets the amplitude at the initial response to unity.
\subsection{Package versions}
\label{sec:org03d5760}
The packages that were part of the session as per \texttt{devtools::session\_info}, are
in Table \ref{tbl:allpkgs}, note that the packages used are only the attached
ones. All packages are from \texttt{CRAN (R 4.3.3)}. The \texttt{R} base was version 4.3.3
(2024-02-29) running on \texttt{Arch Linux (x86\_64, linux-gnu)} with an \texttt{X11 UI},
English language, \texttt{en\_US.UTF-8} collation and \texttt{ctype}, UTC timezone.

\begin{table}[htbp]
\label{tbl:allpkgs}
\centering
\begin{tabular}{lrr}
packages.package & packages.loadedversion & packages.date\\
\hline
broom & 1.0.7 & 2024-09-26\\
changepoint & 2.30 & 2024-11-04\\
cluster & 2.1.8 & 2024-12-11\\
devtools & 2.4.5 & 2022-10-11\\
dplyr & 1.1.4 & 2023-11-17\\
factoextra & 1.0.7 & 2020-04-01\\
forcats & 1.0.0 & 2023-01-29\\
fs & 1.6.5 & 2024-10-30\\
ggnewscale & 0.5.1 & 2025-02-24\\
ggplot2 & 3.5.1 & 2024-04-23\\
ggpmisc & 0.6.1 & 2024-11-14\\
ggpp & 0.5.8-1 & 2024-07-01\\
ggthemes & 5.1.0 & 2024-02-10\\
httpgd & 1.3.1 & 2023-01-30\\
khroma & 1.16.0 & 2025-02-25\\
latex2exp & 0.9.6 & 2022-11-28\\
lubridate & 1.9.4 & 2024-12-08\\
patchwork & 1.3.0 & 2024-09-16\\
pracma & 2.4.4 & 2023-11-10\\
purrr & 1.0.4 & 2025-02-05\\
readr & 2.1.5 & 2024-01-10\\
readxl & 1.4.5 & 2025-03-07\\
reprex & 2.1.1 & 2024-07-06\\
scales & 1.3.0 & 2023-11-28\\
stringr & 1.5.1 & 2023-11-14\\
testthat & 3.2.3 & 2025-01-13\\
tibble & 3.2.1 & 2023-03-20\\
tidyr & 1.3.1 & 2024-01-24\\
tidyverse & 2.0.0 & 2023-02-22\\
usethis & 3.1.0 & 2024-11-26\\
\end{tabular}
\end{table}
\section{Results}
\label{sec:org9fcd33a}
\subsection{Single Beam Z-Scan}
\label{sec:org9ed4ba7}
\begin{figure}[htbp]
\centering
\includegraphics[width=.9\linewidth]{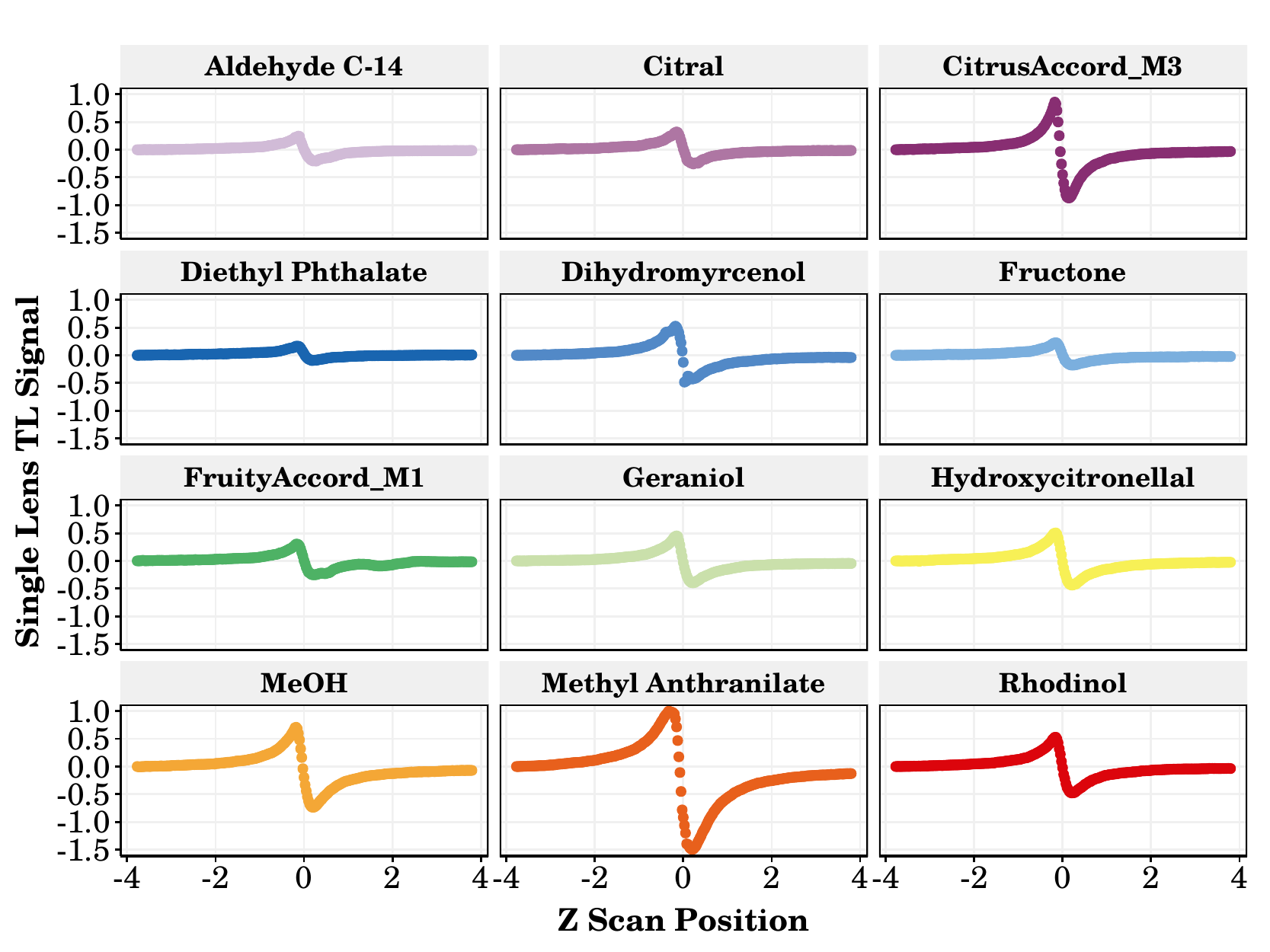}
\caption{\label{fig:sbz_bycomp}Single beam Z-scan data arranged by component.}
\end{figure}

\begin{figure}[htbp]
\centering
\includegraphics[width=.9\linewidth]{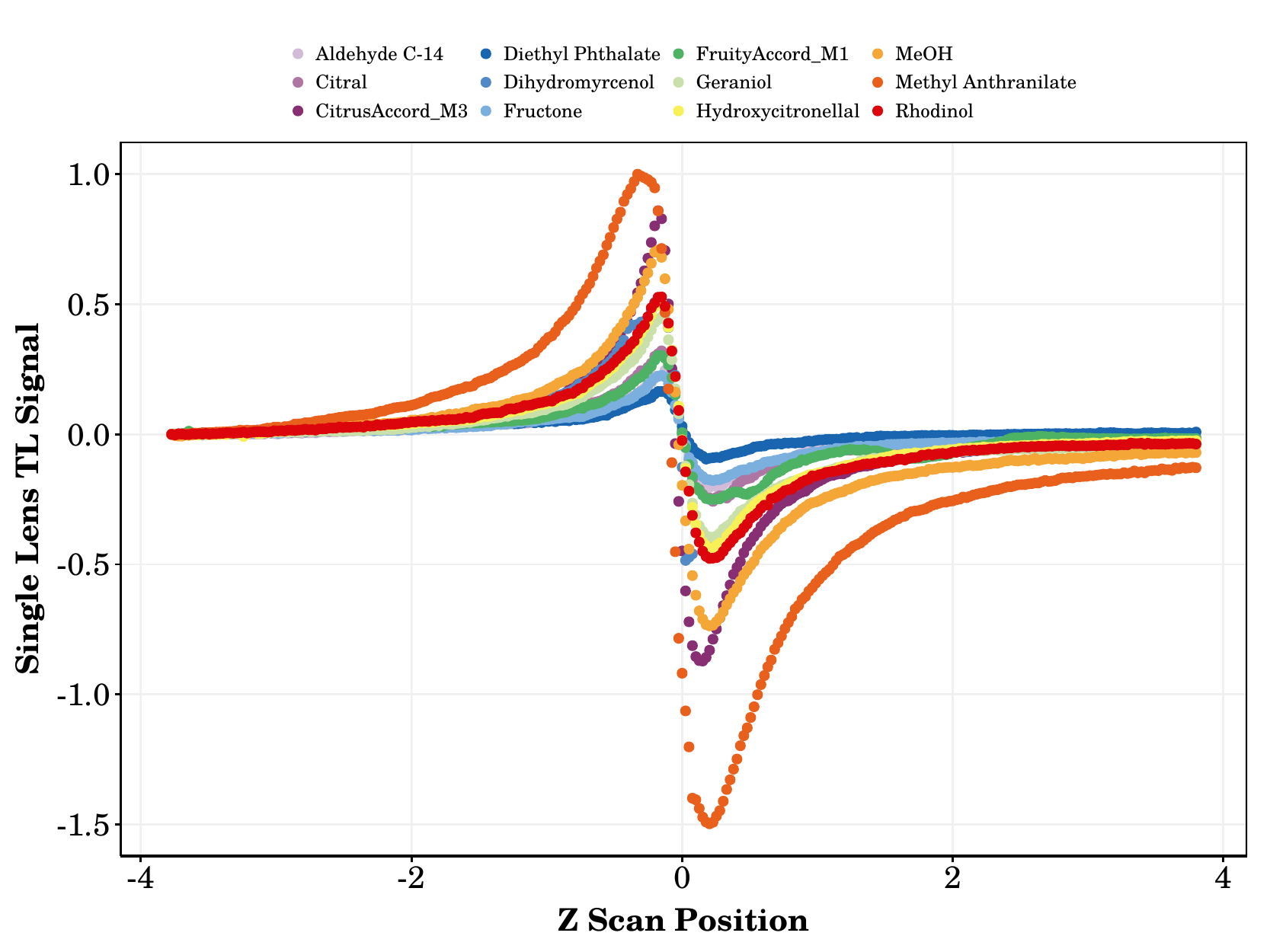}
\caption{\label{fig:sbz_allinone}Single beam Z-scan data in one figure for scale.}
\end{figure}

The raw data for the single beam Z-scan is presented in Figures \ref{fig:sbz_bycomp} and \ref{fig:sbz_allinone}.

\subsubsection{Comparison to Tpv}
\label{sec:orge1bf6e0}
\begin{figure}[htbp]
\centering
\includegraphics[width=.9\linewidth]{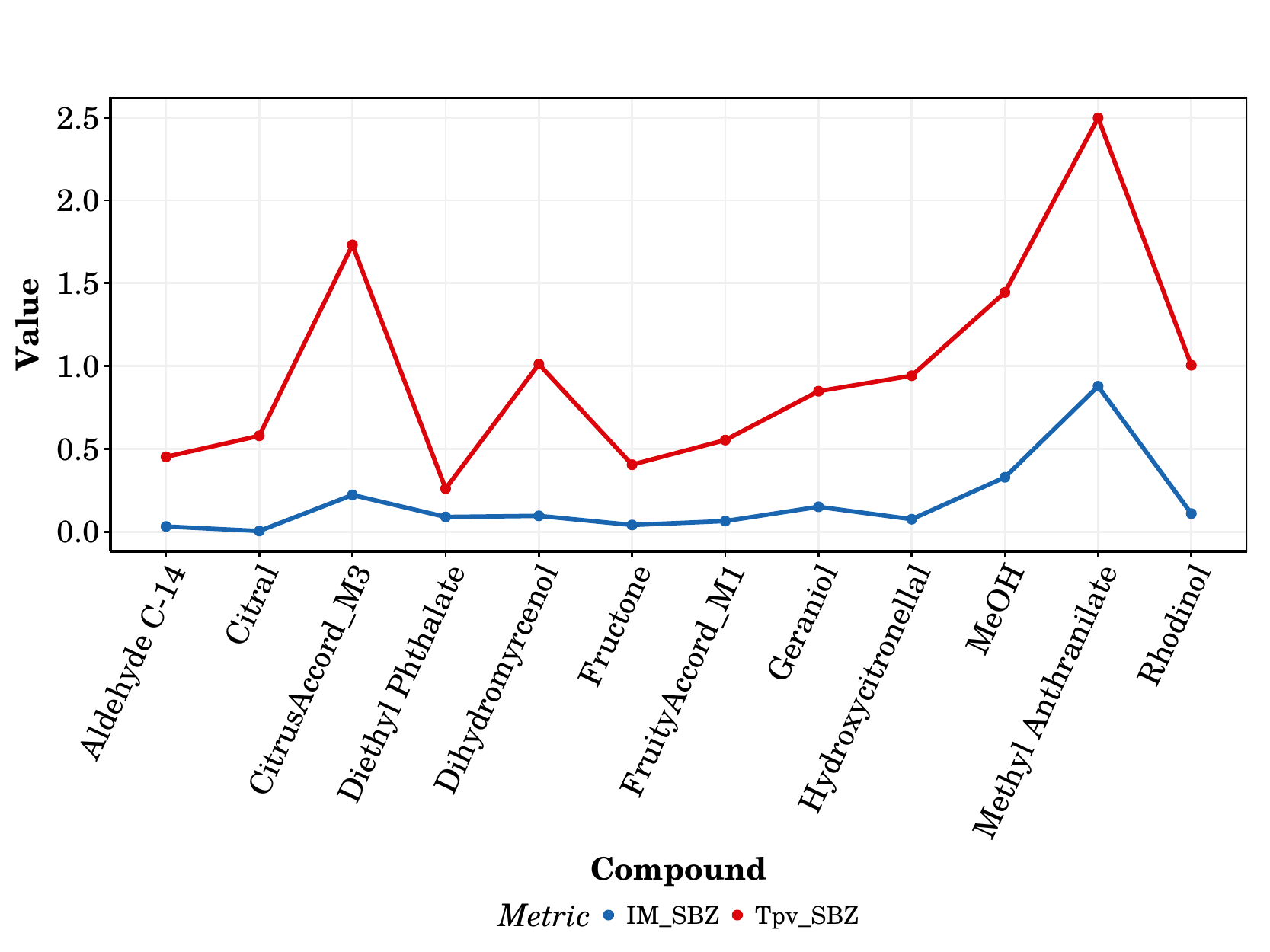}
\caption{\label{fig:sbz_tpv}\(\Delta T_{pv}\) and FTL-IM comparison for the single beam Z-scan data.}
\end{figure}

The peak to valley measure is given by Eq. \ref{eq:tpv_sbz}, while the comparison to the FTL-IM is in Figure \ref{fig:sbz_tpv}.
\begin{equation}
Tpv_{SBZ} = \max(TL_{SBZ}) - \min(TL_{SBZ})
\label{eq:tpv_sbz}
\end{equation}

\subsection{Dual Beam Z-Scan}
\label{sec:orgfafbf37}
\begin{figure}[htbp]
\centering
\includegraphics[width=.9\linewidth]{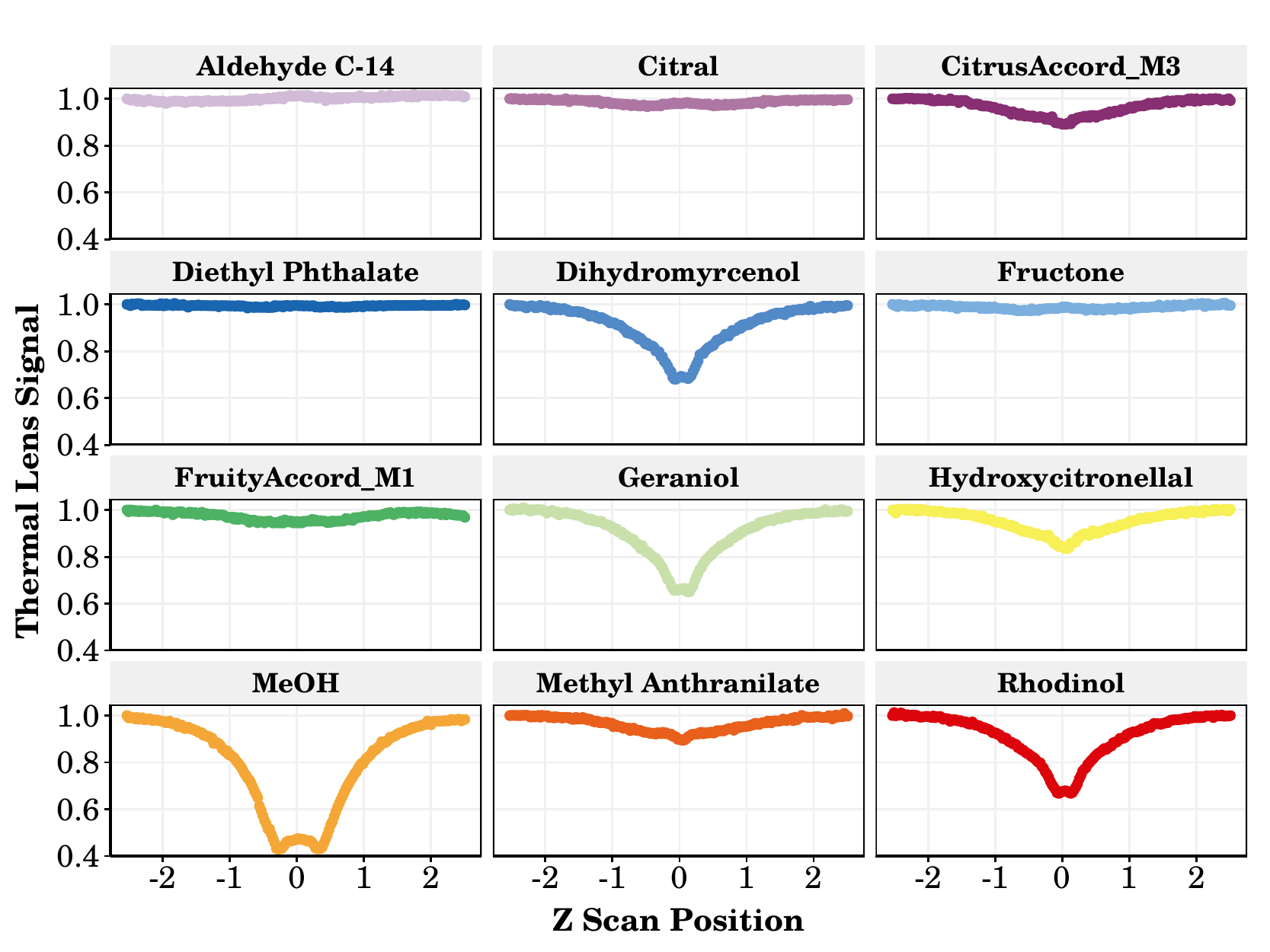}
\caption{\label{fig:dbz_bycomp}Dual beam Z-scan data arranged by component.}
\end{figure}

\begin{figure}[htbp]
\centering
\includegraphics[width=.9\linewidth]{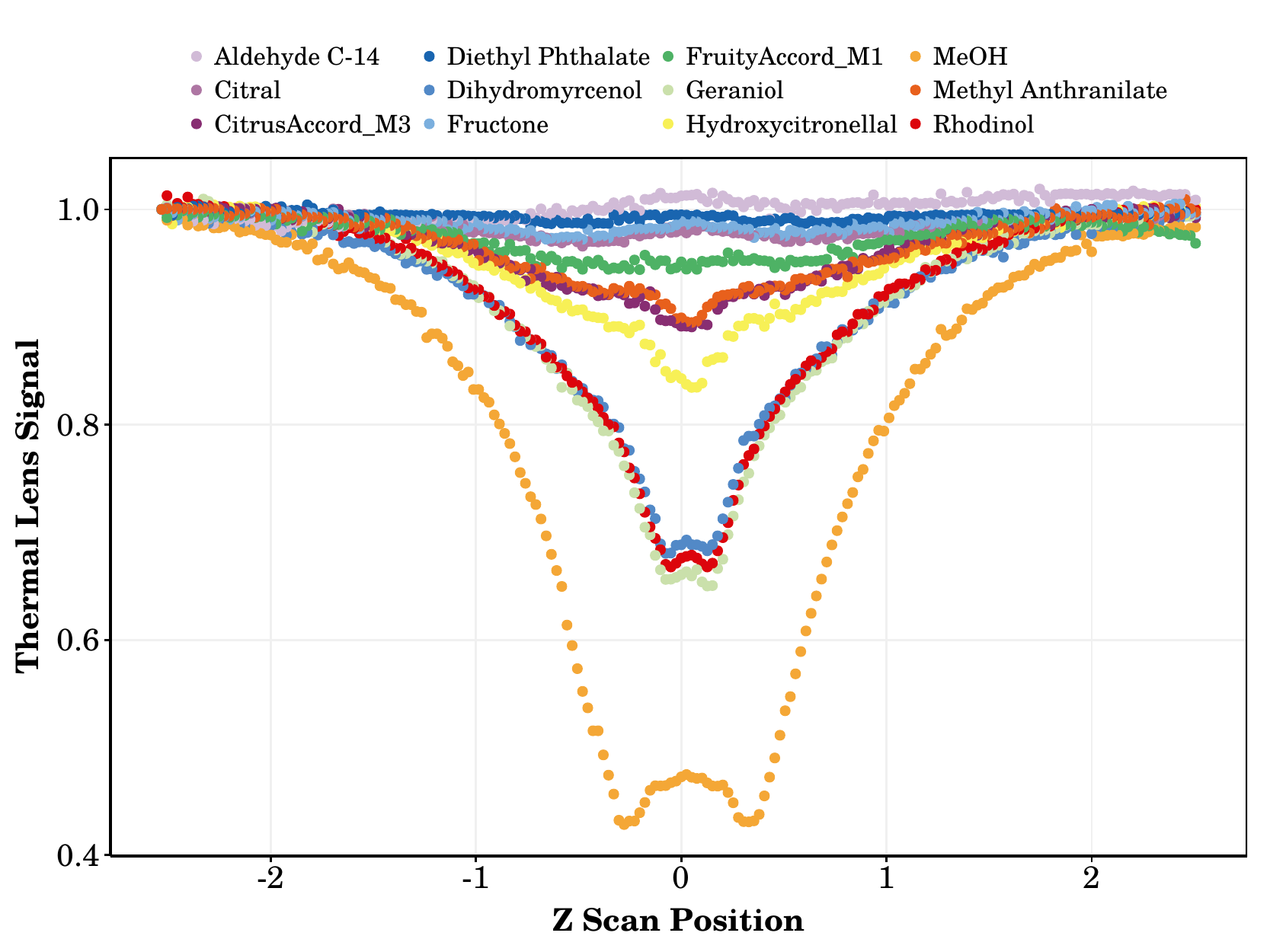}
\caption{\label{fig:dbz_allinone}Dual beam Z-scan data in one figure for scale.}
\end{figure}

The raw data for the dual beam Z-scan is presented in Figures \ref{fig:dbz_bycomp} and \ref{fig:dbz_allinone}.

\subsubsection{Comparison to \(Z=0\) signal}
\label{sec:orgfc2654e}
\begin{figure}[htbp]
\centering
\includegraphics[width=.9\linewidth]{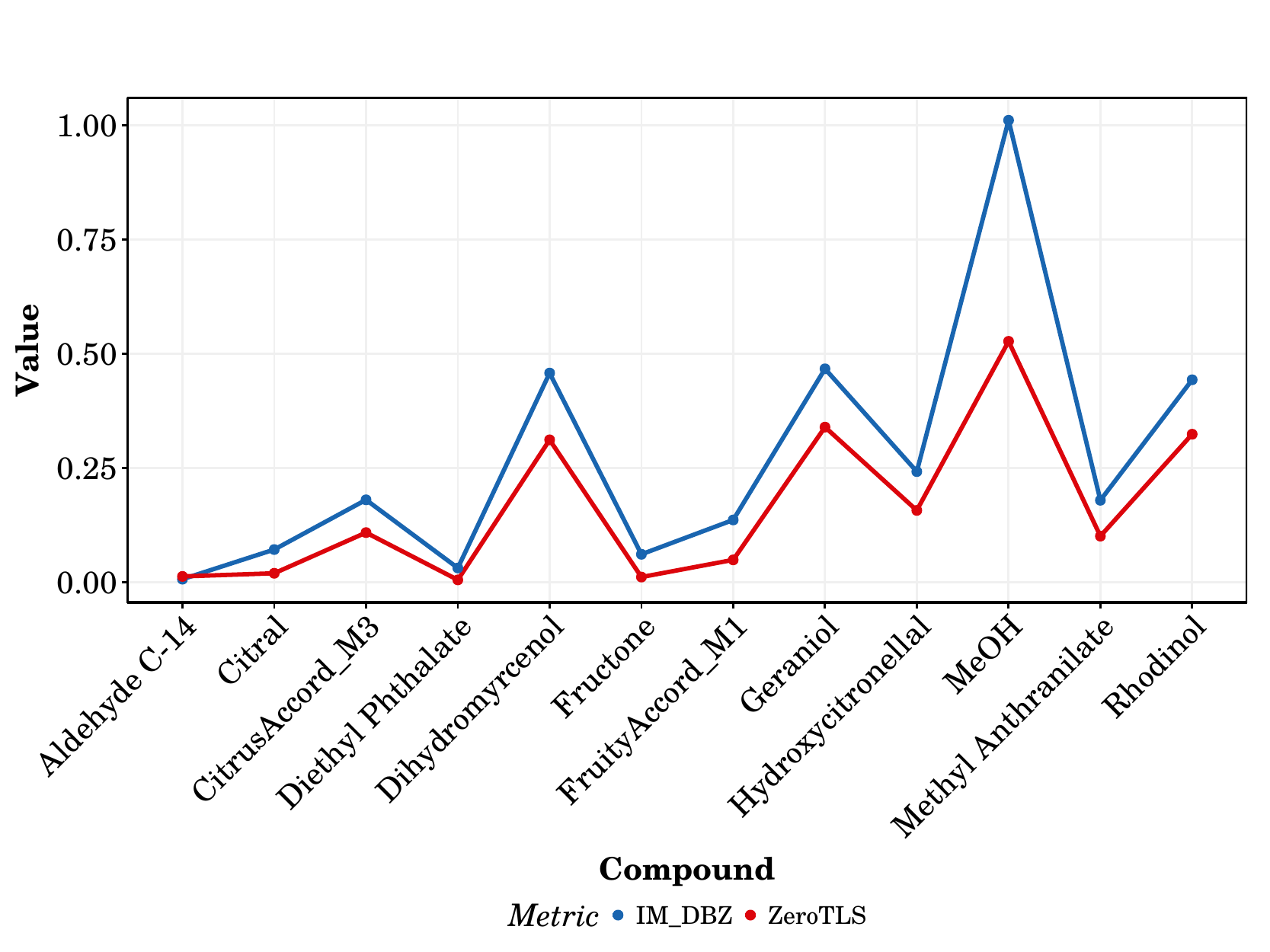}
\caption{\label{fig:dbz_ztls}\(Z==0\) signal and FTL-IM comparison for the dual beam Z-scan data.}
\end{figure}

The signal at \(Z=0\) corresponds to the focal point signal, and is given by Eq. \ref{eq:ztl_dbz}.
\begin{equation}
ZTL_{DBZ} = |1 - TL_{DBZ}(Z=0)|
\label{eq:ztl_dbz}
\end{equation}
Where \(TL_{DBZ}(Z=0)\) represents the value of the baseline-corrected and shifted
thermal lensing signal when the Z-position is equal to \(0\). The comparison
between our measure and this zero point signal is shown in Figure
\ref{fig:dbz_ztls}.

\subsubsection{FTLS-IM Validation}
\label{sec:orgc7bdb9b}
\begin{figure}[htbp]
\centering
\includegraphics[width=.9\linewidth]{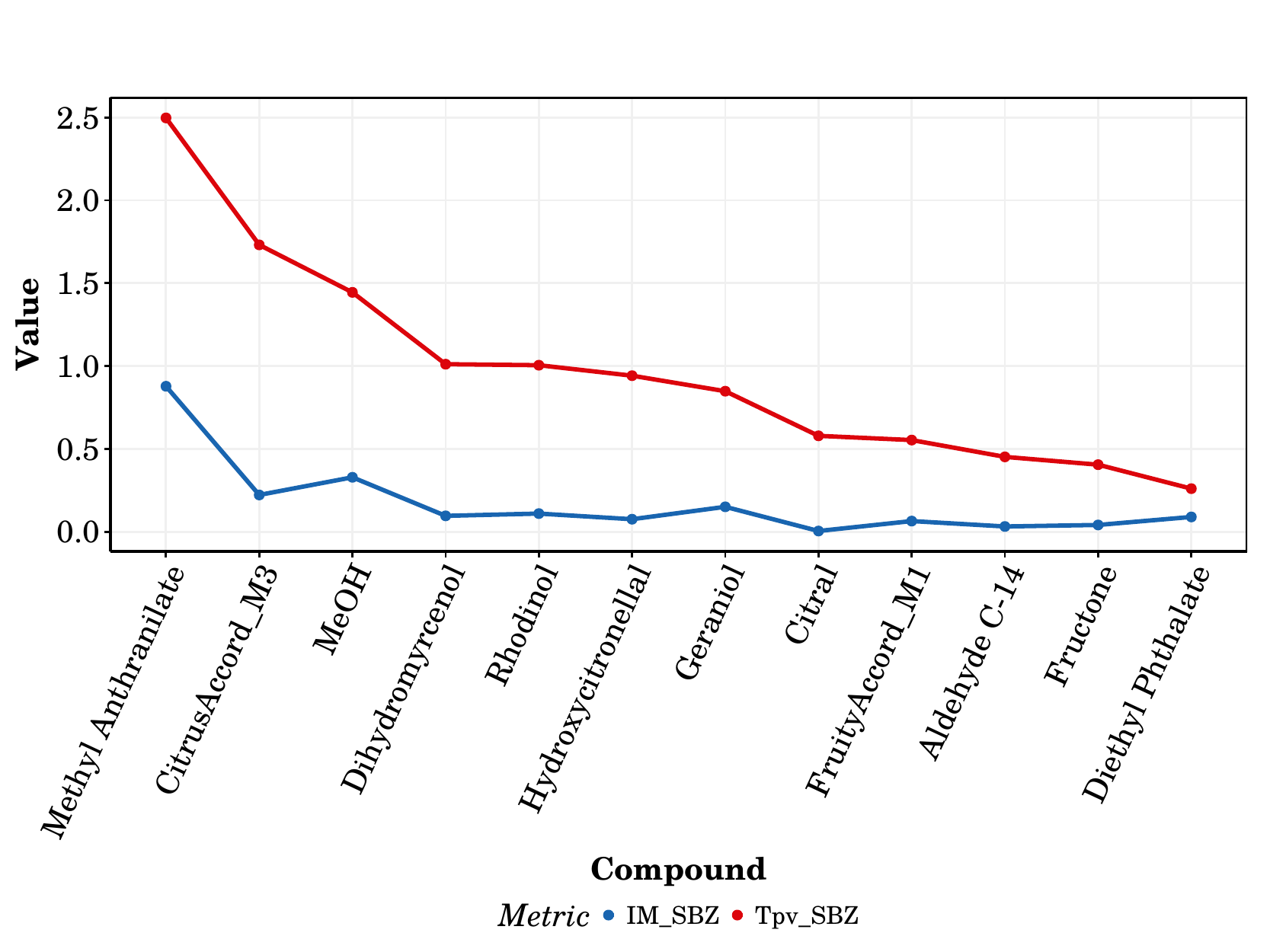}
\caption{\label{fig:ftls_im_zpl}FTL-IM and \(Z=0\) measures for dual beam Z-scan data. All trends are equivalent, though the FTL-IM shows an increased in magnitude for systems with convective dissipation dynamics (``W'' shape).}
\end{figure}

The ``W'' shape effect is captured by the FTL-IM measure as seen in Figure
\ref{fig:ftls_im_zpl}, where it is seen to correspond to a higher measure compared
to the signal measured at \(Z=0\). The trends are equivalent, although the
magnitude of the FTL-IM compared to the single point measure reinforces the
understanding that the measure is sensitive to convective effects. The value for
MeA and Citrus accord in both measures is almost equivalent, and it is also
visually evident that the measure order corresponds exactly to the transmittance
order.

\subsection{Dual Beam Time Resolved}
\label{sec:orge6453c0}
\begin{figure}[htbp]
\centering
\includegraphics[width=.9\linewidth]{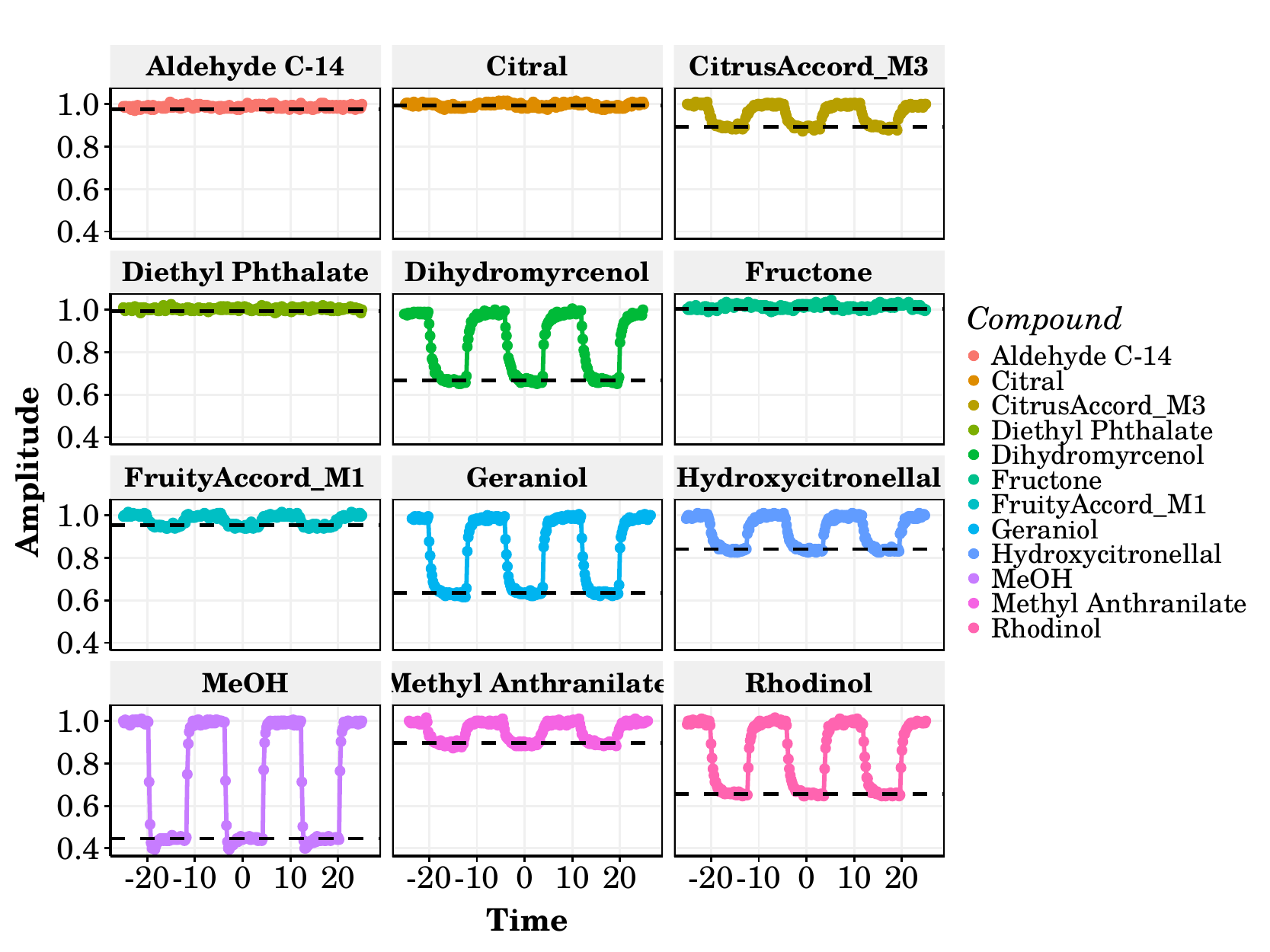}
\caption{\label{fig:dtr_bycomp_sstl}DTR data arranged by component, with SSTL overlaid.}
\end{figure}

\begin{figure}[htbp]
\centering
\includegraphics[width=.9\linewidth]{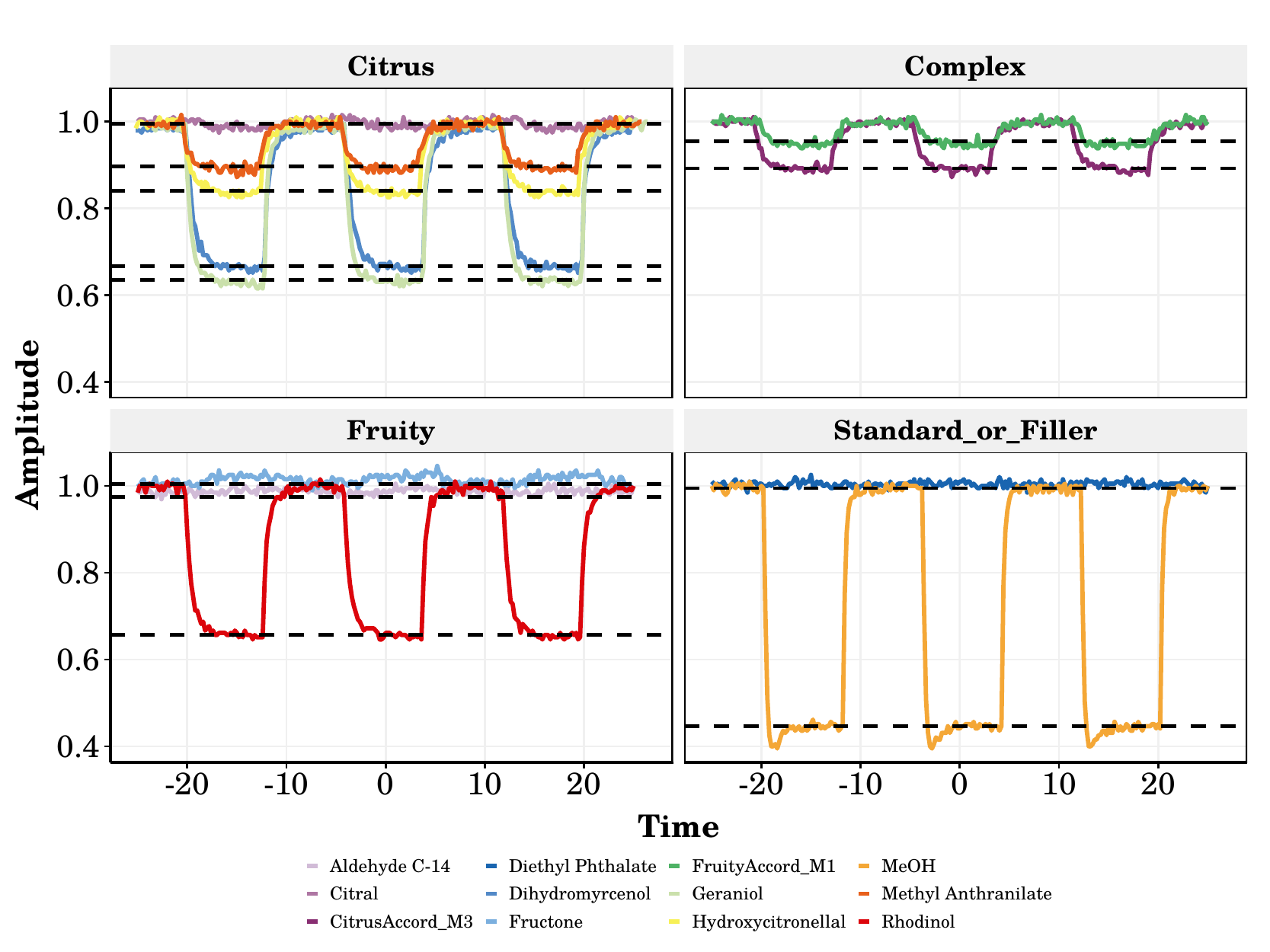}
\caption{\label{fig:dtr_byaccord}DTR data arranged by accord, with SSTL overlaid.}
\end{figure}

The raw data for the dual beam time resolved setup is visualized in Figures
\ref{fig:dtr_bycomp_sstl} and \ref{fig:dtr_byaccord}.

\begin{figure}[htbp]
\centering
\includegraphics[width=.9\linewidth]{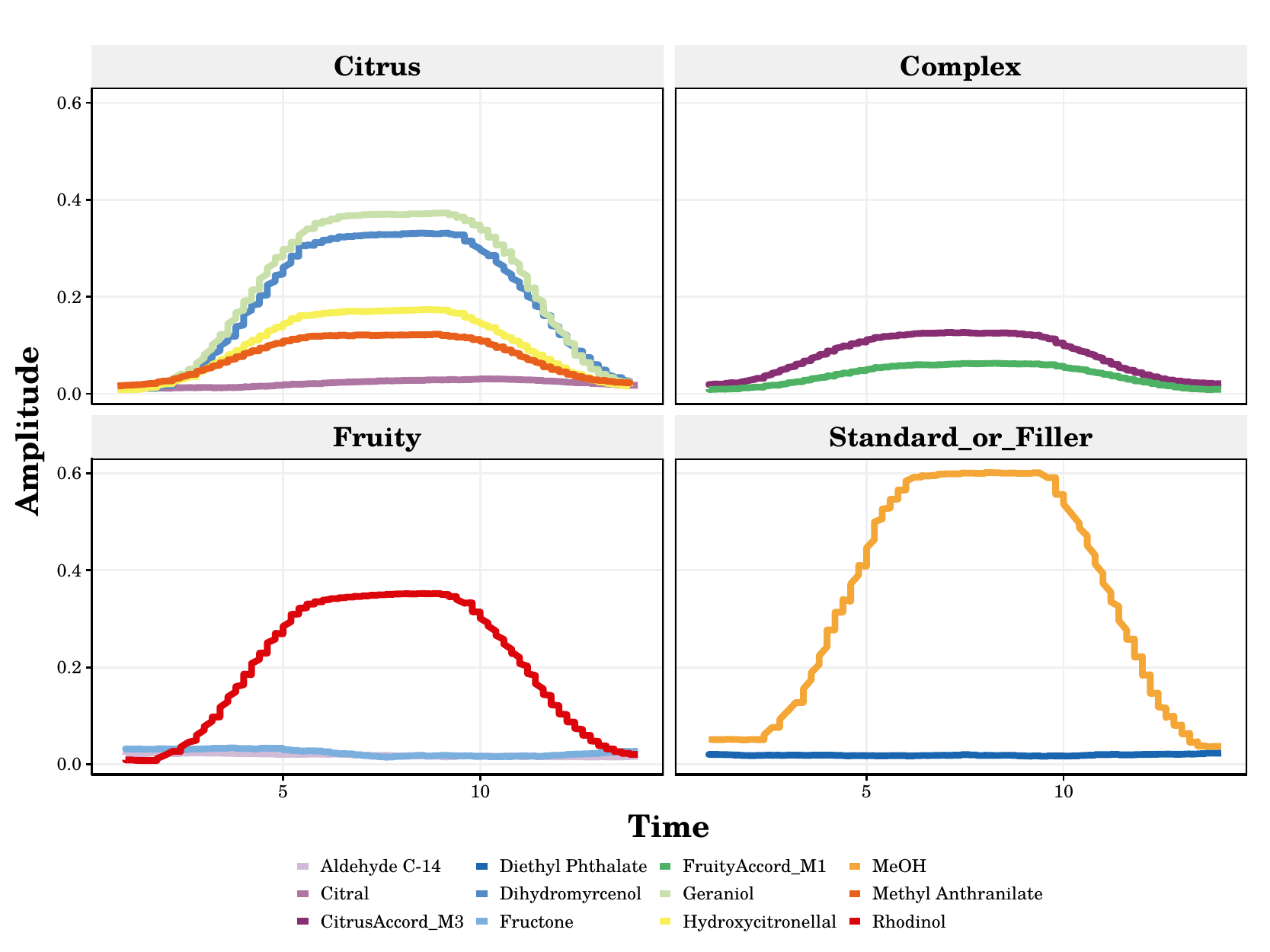}
\caption{\label{fig:dtr_dat_sm}Dual beam Z-scan data arranged by accord. MeOH is the reference point, and DEP shows no appreciable signal.}
\end{figure}

To facilitate visual comparison between compounds and accords,
the data were smoothed using a rolling mean and folded to overlay multiple
cycles (Figure \ref{fig:dtr_dat_sm}).

\subsubsection{Steady state TL determination}
\label{sec:orgcdbea79}
The steady-state thermal lens (SSTL) signal, represents the amplitude of the
thermal lens signal after it has reached equilibrium following the initial
transient response.  Accurately determining this value is crucial for subsequent
calculations, such as the figure of merit (FTL-IM).  We employed a changepoint
detection method to identify the point at which the signal stabilizes.

The procedure involved the following steps:

\begin{description}
\item[{Data Pre-filtering}] For each compound, the time-resolved thermal lens signal
data (\texttt{Amplitude\_shifted}, already baseline-corrected and shifted) was
pre-filtered to consider only the portion of the curve below the mean
amplitude. This pre-filtering step was implemented to prevent the initial,
rapid decrease in signal (often associated with the formation of the thermal
lens) from being incorrectly identified as the steady-state. We focused on the
lower half of the signal where the stabilization occurs.

\item[{Changepoint Analysis}] The \texttt{changepoint} package in \texttt{R} was used to identify
the changepoint within the pre-filtered data. Specifically, the \texttt{cpt.meanvar}
function was applied. This function is designed to detect changes in both the
mean and variance of a time series.  The minimum segment length between
changepoints was set to the \texttt{ESTIMATED\_PERIOD} of \(15\). The \texttt{CROPS}
(Changepoints for a Range Of Penalties) penalty
\cite{haynesEfficientPenaltySearch2014} was employed with the \texttt{pen.value} set
between 0 and 1. Since the changepoints will be, at most, one value, then the
first and only value is taken.

\item[{Steady-State Value Extraction}] The changepoint analysis identified the
index (time point) at which the signal transitioned to its steady-state
behavior.  The STL value was then extracted as the \texttt{Amplitude\_shifted} value
corresponding to that changepoint index.
\end{description}
\subsubsection{FTLS-IM Validation}
\label{sec:org68f3fcd}
\begin{figure}[htbp]
\centering
\includegraphics[width=.9\linewidth]{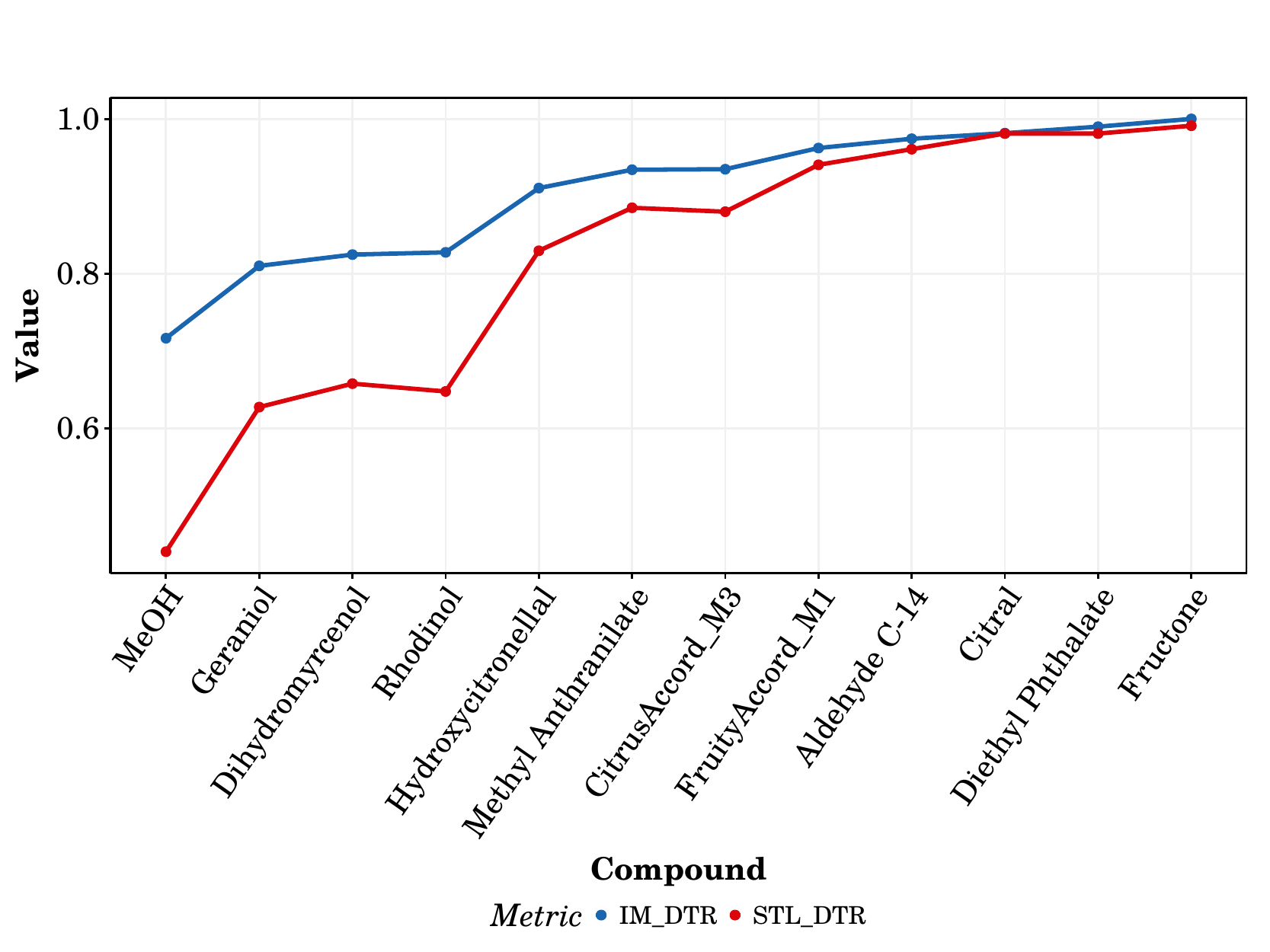}
\caption{\label{fig:ftls_im_stl}FTL-IM and steady state TL measures for dual beam time resolved data. All trends are equivalent, though the FTL-IM shows an increased in magnitude for systems with convective dissipation dynamics (dip below STL).}
\end{figure}

For the time resolved data, the FTL-IM measure is contrasted against the the
steady state TL signal in Figure \ref{fig:ftls_im_stl}, the primary difference is
that of magnitude, as the FTL-IM has a larger value for alcohols.

\begin{figure}[htbp]
\centering
\includegraphics[width=.9\linewidth]{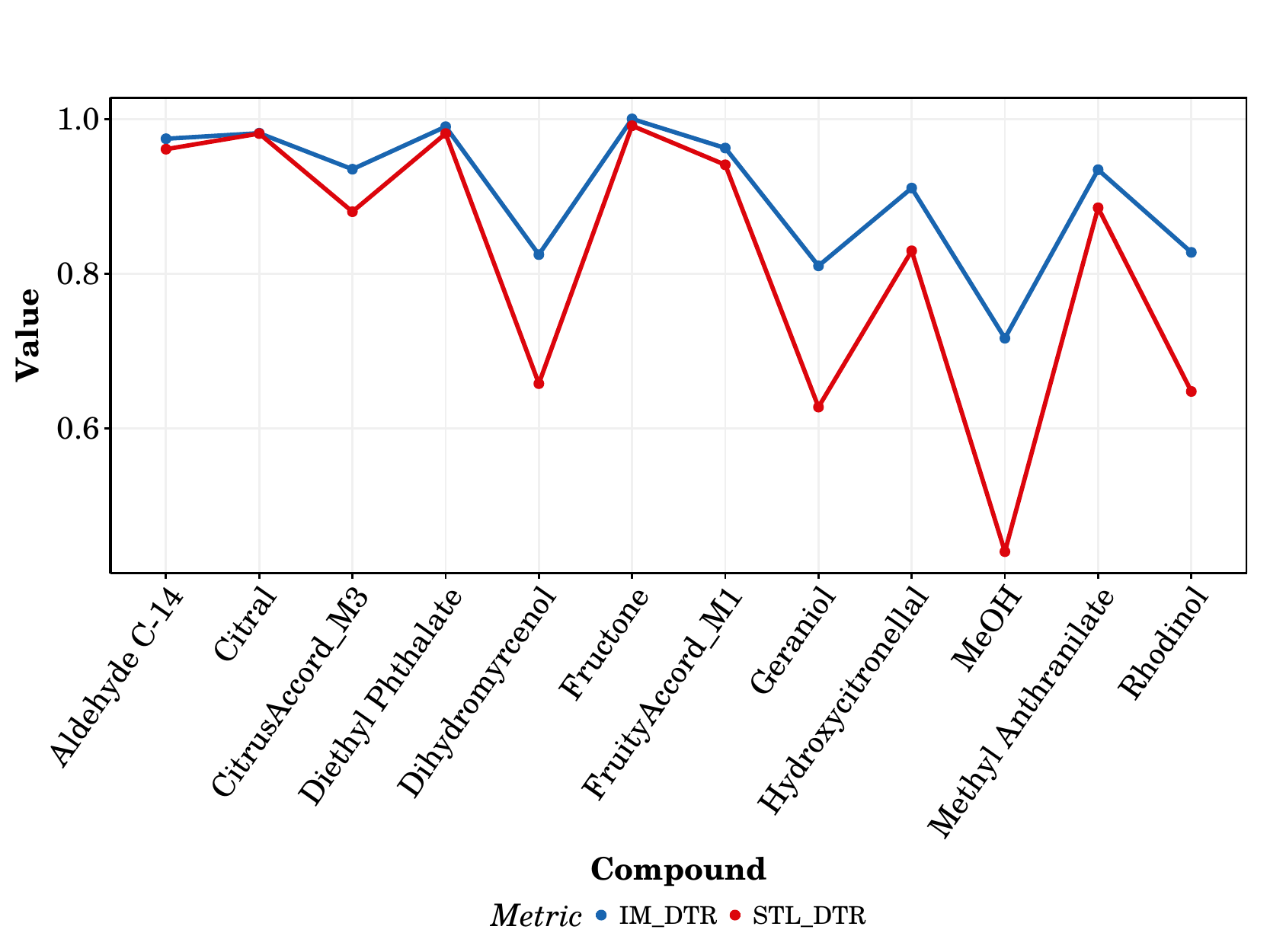}
\caption{\label{fig:dtr_stl}STL for dual beam time resolved data and FTL-IM comparison, without sorting by value.}
\end{figure}

The same data is shown without sorting by either metric in Figure
\ref{fig:dtr_stl}.

\subsection{Comparing FTL-IM}
\label{sec:org4c153a6}
The FTL-IM values for the SBZ and DBZ are both on similar scales by construction. For the DTR result, we scale using min-max scaling given by Eq. \ref{eq:dtr_ms}.
\begin{equation}
IM_{ms, DTR} = \frac{IM_{DTR} - \min(IM_{DTR})}{\max(IM_{DTR}) - \min(IM_{DTR})}
\label{eq:dtr_ms}
\end{equation}

\end{appendices}
\newpage
\bibliography{fragnancePaper,manual_refs}
\end{document}